\documentclass[11pt,preprintnumbers,titlepage,nofootinbib]{revtex4-2}%
\usepackage[latin9]{inputenc}
\usepackage{amsmath}
\usepackage{amssymb}
\usepackage{graphicx}
\usepackage[unicode=true,  bookmarks=false,  breaklinks=false,pdfborder={0 0
1},backref=section,colorlinks=false]{hyperref}
\usepackage{wasysym}
\usepackage{array}
\usepackage{multirow}
\usepackage{wasysym}
\usepackage{wasysym}
\usepackage{amsfonts}
\usepackage{subfigure}
\usepackage{cleveref}
\usepackage{listings}
\usepackage{xcolor}
\usepackage{array}
\usepackage{url}
\usepackage{color}
\usepackage{afterpage}
\usepackage{float}%
\hypersetup{
colorlinks,linkcolor=blue,citecolor=blue,urlcolor=blue}
\makeatletter

\@ifundefined{textcolor}{}{
\definecolor{BLACK}{gray}{0}
\definecolor{WHITE}{gray}{1}
\definecolor{RED}{rgb}{1,0,0}
\definecolor{GREEN}{rgb}{0,1,0}
\definecolor{BLUE}{rgb}{0,0,1}
\definecolor{CYAN}{cmyk}{1,0,0,0}
\definecolor{MAGENTA}{cmyk}{0,1,0,0}
\definecolor{YELLOW}{cmyk}{0,0,1,0}
}

\makeatother
\begin{document}
\preprint{CTP-SCU/2024009}
\title{Stationary Scalar Clouds around Kerr-Newman Black Holes}
\author{Guangzhou Guo$^{a,b}$}
\email{guangzhouguo@outlook.com}
\author{Peng Wang$^{a}$}
\email{pengw@scu.edu.cn}
\author{Tianshu Wu$^{a}$}
\email{wutianshu@stu.scu.edu.cn}
\author{Haitang Yang$^{a}$}
\email{hyanga@scu.edu.cn}
\affiliation{$^{a}$Center for Theoretical Physics, College of Physics, Sichuan University,
Chengdu, 610064, China}
\affiliation{$^{b}$Department of Physics, Southern University of Science and Technology,
Shenzhen, 518055, China}

\begin{abstract}
This study investigates scalar clouds around Kerr-Newman black holes within
the Einstein-Maxwell-scalar model. Tachyonic instabilities are identified as
the driving mechanism for scalar cloud formation. Employing the spectral
method, we numerically compute wave functions and parameter space existence
domains for both fundamental and excited scalar cloud modes. Our analysis
demonstrates that black hole spin imposes an upper limit on the existence of
scalar clouds, with excited modes requiring stronger tachyonic instabilities
for their formation. These findings lay the groundwork for exploring the
nonlinear dynamics and astrophysical implications of scalar clouds.

\end{abstract}
\maketitle
\tableofcontents

{}

{}

{}

\section{Introduction}

In the context of electro-vacuum general relativity, the no-hair theorem
establishes that all stationary black holes are uniquely defined by their
mass, angular momentum and electric charge
\cite{Israel:1967wq,Carter:1971zc,Ruffini:1971bza,Wu:2024ixf,Chen:2023uuy}.
This theorem not only addresses the existence of stationary black holes with
more degrees of freedom but also describes the dynamical endpoint of
gravitational collapse. Testing the no-hair theorem is crucial for
understanding black hole physics and can help constrain alternative theories
of gravity. For instance, the black-hole spectroscopy program, which analyzes
quasinormal modes extracted from gravitational-wave observations, has emerged
as a powerful tool for probing the Kerr nature of astrophysical compact
objects \cite{Isi:2019aib,Bhagwat:2019dtm,Wang:2021elt}.

While the no-hair theorem dictates that stationary black holes in
electro-vacuum spacetime are uniquely described by mass, angular momentum and
electric charge, the discovery of the first hairy black hole solution within
Einstein-Yang-Mills theory challenged this notion
\cite{Volkov:1989fi,Bizon:1990sr,Greene:1992fw}. Since then, numerous
counterexamples have emerged
\cite{Luckock:1986tr,Droz:1991cx,Kanti:1995vq,Sotiriou:2013qea,Cisterna:2014nua,Antoniou:2017acq}%
. In particular, black holes with scalar hair have attracted great interest,
as scalar fields are well-motivated beyond the standard model and can be used
to model dark energy and dark matter \cite{Herdeiro:2015waa}. However, the
existence of asymptotically flat black holes with scalar hair is constrained
by the Bekenstein's uniqueness theorem for $V$-scalar-vacuum
\cite{Bekenstein:1971hc,Bekenstein:1972ny,Bekenstein:1972ky}. Therefore, such
black holes require specific conditions: a non-minimal coupling of the scalar
field
\cite{Herdeiro:2018wub,Doneva:2018rou,Cunha:2019dwb,Herdeiro:2020wei,Berti:2020kgk}%
, a violation of spacetime symmetry for the scalar field
\cite{Herdeiro:2014goa,Wang:2018xhw,Brihaye:2021mqk}, or a scalar potential
$V$ that deviates from certain limitations
\cite{Nucamendi:1995ex,Gubser:2005ih,Kleihaus:2013tba}.

In the absence of a significant backreaction from the scalar field, hairy
black holes can be approximated as their bald counterparts enveloped by a
scalar cloud. This cloud represents a bound state solution of the scalar field
within the bald black hole's spacetime. Scalar clouds often signal the
emergence of hairy black holes and serve as seeds for numerical searches of
hairy black hole solutions. Consequently, a thorough understanding of their
existence offers invaluable insights into the formation mechanisms of hairy
black holes. Furthermore, the unique signatures of scalar clouds, particularly
those composed of axion-like particles, have been leveraged to impose
stringent constraints on the scalar field's parameter space. These constraints
hold significant implications for dark matter exploration and research into
physics beyond the Standard Model
\cite{Arvanitaki:2009fg,Brito:2017wnc,Davoudiasl:2019nlo,Chen:2019fsq,Chen:2022kzv,Chen:2023vkq}%
.

Sustaining scalar clouds outside black holes necessitates a mechanism to
counteract the inward flow of the scalar field across the event horizon. A
prominent example is superradiance, which can extract energy from rotating or
charged black holes \cite{Brito:2015oca}. Under superradiant conditions, the
existence of scalar clouds surrounding stationary, rotating black holes has
been established for complex scalar fields
\cite{Hod:2014baa,Benone:2014ssa,Huang:2017whw,Kunz:2019bhm,Santos:2021nbf}.
Note that these clouds exhibit a phase-like time dependence, violating an
assumption of the Bekenstein's theorem. However, their energy-momentum tensor
remains time-independent, resulting in stationary black holes with
synchronized hair \cite{Delgado:2020hwr}.

Alternatively, non-minimal couplings between scalar fields and curvature
invariants have been shown to induce tachyonic instabilities in the scalar
fields \cite{Cardoso:2013opa,Cardoso:2013fwa}. These instabilities can trigger
exponential growth of the scalar field, counteracting the leakage through the
event horizon. Consequently, sufficiently strong tachyonic instabilities can
lead to the formation of scalar clouds, potentially serving as a threshold for
the emergence of hairy black holes
\cite{Doneva:2017bvd,Silva:2017uqg,Antoniou:2017acq,Cunha:2019dwb,Cunha:2019dwb,Guo:2021zed,Xu:2024cfe}%
. This phenomenon, known as ``spontaneous scalarization,''\ endows general
relativistic stars and black holes with a non-trivial scalar configuration
only above a certain threshold of spacetime curvature \cite{Damour:1993hw}.
Therefore, spontaneous scalarization allows scalarized compact objects to
acquire a non-trivial scalar configuration solely in regimes of strong
gravity, enabling them to evade constraints derived from weak-field gravity tests.

Recent studies have demonstrated the existence of scalarized
Reissner-Nordstr\"{o}m (RN) black holes within specific
Einstein-Maxwell-scalar (EMS) models featuring non-minimal couplings between
the scalar and Maxwell fields \cite{Herdeiro:2018wub}. Notably, the existence
of these black holes is bounded by scalar clouds and critical lines within the
parameter space. Interestingly, for certain parameter regimes, scalarized RN
black holes have been found to possess two photon spheres outside the event
horizon \cite{Gan:2021pwu}. This unique feature leads to distinct
phenomenology, including black hole images with intricate structures
\cite{Gan:2021xdl,Guo:2022muy,Chen:2022qrw,Chen:2023qic,Chen:2024ilc} and echo
signals \cite{Guo:2021enm,Guo:2022umh}. Furthermore, investigations into
superradiant instabilities and non-linear stability of these double photon
sphere black holes have been conducted \cite{Guo:2023ivz,Guo:2024cts}. For a
comprehensive analysis of black holes with multiple photon spheres, we refer
readers to \cite{Guo:2022ghl}.

Building upon the work of \cite{Herdeiro:2018wub}, we extended the analysis to
rotating charged black holes in \cite{Guo:2023mda}. Similar to the case of RN
black holes, tachyonic instabilities were found to induce scalar cloud
formation around Kerr-Newman (KN) black holes, resulting in scalarized KN
black holes within the EMS model. Interestingly, within specific parameter
spaces, these scalarized KN black holes were shown to possess two unstable and
one stable light ring on the equatorial plane, for both prograde and
retrograde directions. However, our previous work in \cite{Guo:2023mda} only
considered the fundamental state for a limited range of non-minimal coupling
values. A more comprehensive exploration of scalar clouds is necessary for a
deeper understanding of tachyonic instabilities and spontaneous scalarization
in KN black holes.

This paper addresses the existing gap by conducting a comprehensive analysis
of fundamental and excited scalar clouds surrounding KN black holes within the
EMS model. The paper is structured as follows. Sec. \ref{sec:EMS Model}
introduces the EMS model and its associated scalar clouds. Sec.
\ref{sec:III. Numerical Framework} outlines the numerical methodology,
employing the spectral method to obtain scalar cloud solutions. Sec.
\ref{sec:Numerical Results} presents and analyzes numerical findings. Finally,
Sec. \ref{sec:Conclusion} summarizes key results and discusses their
implications. Throughout this paper, we adopt units where $G=c=4\pi
\epsilon_{0}=1$.

\section{Einstein-Maxwell-scalar Model}

\label{sec:EMS Model}

The action of the EMS model is given by
\begin{equation}
S=\frac{1}{16\pi}\int d^{4}x\sqrt{-g}\left[  R-2\partial_{\mu}\Phi
\partial^{\mu}\Phi-f\left(  \Phi\right)  F^{\mu\nu}F_{\mu\nu}\right]
,\label{eq:Action}%
\end{equation}
where $A_{\mu}$ represents the electromagnetic field, $\Phi$ is the scalar
field, and $F_{\mu\nu}=\partial_{\mu}A_{\nu}-\partial_{\nu}A_{\mu}$ denotes
the electromagnetic field strength tensor. In the EMS model, the scalar field
is non-minimally coupled to electromagnetism via the the coupling function
$f\left(  \Phi\right)  $. The scalar field equation of motion is%
\begin{equation}
\square\Phi=f^{\prime}\left(  \Phi\right)  F_{\mu\nu}F^{\mu\nu}%
/4,\label{eq:seq}%
\end{equation}
indicating that the existence of a scalar-free solution with $\Phi=0$ requires
$f^{\prime}\left(  0\right)  \equiv\left.  df\left(  \Phi\right)
/d\Phi\right\vert _{\Phi=0}=0$. Without loss of generality, we assume
$f\left(  0\right)  =1$. Consequently, at $\Phi=0$, the coupling function
$f\left(  \Phi\right)  $ can be expanded as
\begin{equation}
f\left(  \Phi\right)  =1+\alpha\Phi^{2}+\mathcal{O}\left(  \Phi^{3}\right)
,\label{eq:fphi}%
\end{equation}
where $\alpha$ represents a dimensionless coupling constant.

The non-minimally coupled scalar field destabilizes the background spacetime
through tachyonic instabilities, leading to spontaneous scalarization in black
holes \cite{Herdeiro:2018wub,Guo:2023mda}. The formation of scalar clouds
around scalar-free black holes signifies the onset of this process.
Linearizing Eq. $\left(  \ref{eq:seq}\right)  $ yields the equation of motion
governing the wave function of the scalar cloud on the scalar-free background
spacetime,
\begin{equation}
\left(  \square-\mu_{\text{eff}}^{2}\right)  \Phi=0, \label{eq:delta phi}%
\end{equation}
where $\mu_{\text{eff}}^{2}=$ $f^{\prime\prime}\left(  0\right)  F_{\mu\nu
}F^{\mu\nu}/4$ represents the effective mass. Tachyonic instabilities arise
when $\mu_{\text{eff}}^{2}<0$, potentially triggering the formation of scalar
clouds. Typically, considering only the leading and quadratic terms in Eq.
$\left(  \ref{eq:fphi}\right)  $ suffices for analyzing the onset of
spontaneous scalarization
\cite{Herdeiro:2018wub,Fernandes:2019rez,Hod:2020ljo}. Consequently, we
neglect self-interactions of the scalar field in this work, resulting in
$\mu_{\text{eff}}^{2}=\alpha F^{\mu\nu}F_{\mu\nu}/2$.

The scalar-free black hole solution within the EMS model is a KN black hole,
expressed in the Boyer-Lindquist coordinates as%
\begin{align}
ds^{2}  &  =-\frac{\triangle}{\Sigma}\left(  dt-a\text{sin}^{2}\theta
d\varphi\right)  ^{2}+\frac{\text{sin}^{2}\theta}{\Sigma}\left[  \left(
r^{2}+a^{2}\right)  d\varphi-adt\right]  ^{2}+\frac{\Sigma}{\triangle}%
dr^{2}+\Sigma d\theta^{2},\nonumber\\
A  &  =Qr\frac{dt-a\text{sin}^{2}\theta d\varphi}{\Sigma},
\end{align}
where\textbf{ }%
\begin{align}
\Sigma &  =r^{2}+a^{2}\cos^{2}\theta,\nonumber\\
\triangle &  =r^{2}-2Mr+a^{2}+Q^{2}.
\end{align}
Here, $Q$ is the black hole charge, and $a$ represents the ratio of black hole
angular momentum $J$ to mass $M$ (i.e., $a\equiv J/M$). The event and Cauchy
horizons are located at the roots of\textbf{ $\triangle$, }given by
$r_{+}=M+\sqrt{M^{2}-a^{2}-Q^{2}}$ and $r_{-}=M-\sqrt{M^{2}-a^{2}-Q^{2}}$,
respectively. For future reference, we introduce the dimensionless reduced
black hole charge and spin, denoted as,%
\begin{equation}
q\equiv Q/M\text{, }\chi\equiv a/M\text{.}%
\end{equation}
In the KN black hole background, the effective mass square from Eq. $\left(
\ref{eq:delta phi}\right)  $ becomes%
\begin{equation}
\mu_{\text{eff}}^{2}=-\frac{\alpha q^{2}\left(  \tilde{r}^{4}-6\chi^{2}%
\tilde{r}^{2}\cos^{2}\theta+\chi^{4}\cos^{4}\theta\right)  M^{2}}{\left(
\tilde{r}^{2}+\chi^{2}\cos^{2}\theta\right)  ^{4}}, \label{eq:negmass}%
\end{equation}
where $\tilde{r}\equiv r/M$. It has been demonstrated that the region where
$\mu_{\text{eff}}^{2}<0$ exists outside the event horizon of KN black holes,
indicating the possibility of scalar cloud formation via tachyonic
instabilities \cite{Guo:2023mda}. Moreover, the spatial extent of the negative
effective mass squared region decreases with increasing black hole spin,
suggesting a potential suppression of scalar clouds for rapidly rotating black holes.

To solve Eq. $\left(  \ref{eq:delta phi}\right)  $ for the scalar field $\Phi
$, we employ a Fourier decomposition in terms of frequency $\omega$ and
azimuthal number $m$,%
\begin{equation}
\Phi\left(  t,r,\theta,\varphi\right)  =\int\frac{d\omega}{2\pi}e^{-i\omega
t}\sum\limits_{m}e^{im\varphi}\tilde{\Phi}\left(  \omega,r,\theta,m\right)  .
\end{equation}
For specified $\omega$ and $m$, Eq. $\left(  \ref{eq:delta phi}\right)  $
reduces to a Partial Differential Equation (PDE) for $\tilde{\Phi}\left(
\omega,r,\theta,m\right)  $ with respect to $r$ and $\theta$. Given that
scalar clouds typically serve as seeds for constructing axisymmetric hairy
black hole solutions, this work primarily considers stationary, axisymmetric
scalar clouds, setting $\omega=m=0$. For brevity, we denote $\tilde{\Phi
}\left(  0,r,\theta,0\right)  $ by $\phi\left(  r,\theta\right)  $ in
subsequent discussions.

In KN black holes, $\phi\left(  r,\theta\right)  $ can be further decomposed
as \cite{Hod:2014baa,Herdeiro:2014goa}
\begin{equation}
\phi\left(  r,\theta\right)  =%
%TCIMACRO{\dsum \limits_{n,l}}%
%BeginExpansion
{\displaystyle\sum\limits_{n,l}}
%EndExpansion
R_{nl}\left(  r\right)  S_{l0}\left(  \theta\right)  , \label{eq:phiexpansion}%
\end{equation}
where $S_{l0}\left(  \theta\right)  $ are spheroidal harmonics, and the radial
function $R_{nl}$ satisfies the radial Teukolsky equation
\cite{Teukolsky:1972my,Teukolsky:1973ha,Teukolsky:1974yv}. Analogous to
hydrogen atoms, the wave function $\phi\left(  r,\theta\right)  $ can be
characterized by a discrete set of numbers $\left(  n,l\right)  $, where
$n=0,1,2\cdots$ is the principal quantum number, and $l=0,1,2\cdots$ is the
angular momentum quantum number. The values of $n$ and $l$ correspond to the
number of nodes of the wave function in the radial and angular directions,
respectively. Specifically, the $\left(  n,l\right)  =\left(  0,0\right)  $
scalar cloud represents the fundamental mode, whose presence often signals the
existence of scalarized black hole solutions. It has been shown that the
fundamental mode corresponds to bifurcation points in the parameter space,
marking the onset of scalarized KN black holes \cite{Guo:2023mda}.

The wave equation for $\phi\left(  r,\theta\right)  $ is separable in KN black
holes, enabling the reduction of the PDE for $\phi\left(  r,\theta\right)  $
to ordinary differential equations for $R_{nl}\left(  r\right)  $. However,
instead of utilizing the method of separation of variables, this paper employs
the spectral method to numerically solve the PDE for $\phi\left(
r,\theta\right)  $. It is important to note that the spectral method does not
necessitate the separability of the scalar wave equation. This constitutes a
significant advantage as the method of separation of variables may prove
inapplicable for computations of scalar clouds surrounding black holes in
frameworks beyond general relativity.

To determine the wave function $\phi\left(  r,\theta\right)  $, appropriate
boundary conditions must be imposed at the event horizon and spatial infinity.
Given the regularity of $\phi\left(  r,\theta\right)  $ across the event
horizon, it is possible to expand $\phi\left(  r,\theta\right)  $ in a series
about $r=r_{+}$,%
\begin{equation}
\phi\left(  r,\theta\right)  =\phi_{0}\left(  \theta\right)  +\left(
r-r_{+}\right)  \phi_{1}\left(  \theta\right)  +\cdots.\label{eq:phiexp}%
\end{equation}
Furthermore, the condition of asymptotic flatness requires that $\phi\left(
r,\theta\right)  $ vanishes as $r$ approaches infinity,%
\begin{equation}
\lim_{r\rightarrow\infty}\phi\left(  r,\theta\right)  =0\text{.}%
\end{equation}
Additionally, axial symmetry, combined with regularity on the symmetry axis,
imposes,%
\begin{equation}
\partial_{\theta}\phi\left(  r,\theta\right)  =0\text{, at }\theta=0\text{ and
}\pi\text{.}%
\end{equation}
These boundary conditions uniquely specify a discrete set of KN black holes
capable of supporting scalar clouds, which corresponds to bifurcation points
in the parameter space. In essence, solving the PDE for $\phi\left(
r,\theta\right)  $ reduces to calculating the eigenvalues and eigenfunctions
of a boundary value problem. The eigenfunctions yield $\phi\left(
r,\theta\right)  $, while the eigenvalues establish a relationship between the
KN black hole parameters.

\section{Numerical Setup}

\label{sec:III. Numerical Framework}

In this paper, we employ spectral methods to numerically solve the wave
equation for $\phi\left(  r,\theta\right)  $. Spectral methods constitute a
well-established approach for solving PDEs \cite{boyd2001chebyshev}. These
methods approximate the exact solution via a finite linear combination of
basis functions. Notably, spectral methods exhibit exponential convergence for
well-behaved functions as the number of degrees of freedom increases,
surpassing the linear or polynomial convergence rates of finite difference or
finite element methods. Recent studies have successfully applied spectral
methods to the search for black hole solutions
\cite{Fernandes:2022gde,Lai:2023gwe,Burrage:2023zvk} and the computation of
black hole quasinormal modes
\cite{Jansen:2017oag,Gan:2019jac,Chung:2023zdq,Chung:2023wkd,Chung:2024ira}.
For a comprehensive overview of spectral methods in the context of black hole
physics, interested readers are referred to \cite{Fernandes:2022gde}.

For numerical implementation, a new radial coordinate is introduced,
\begin{equation}
x=\frac{\sqrt{r^{2}-r_{+}^{2}}-r_{+}}{\sqrt{r^{2}-r_{+}^{2}}+r_{+}}.
\end{equation}
This mapping transforms the event horizon and spatial infinity to $x=-1$ and
$x=1$, respectively. Consequently, the expansion in Eq. $\left(
\ref{eq:phiexp}\right)  $ becomes a series expansion of $\phi\left(
x,\theta\right)  $ at $x=-1$,%
\begin{equation}
\phi\left(  x,\theta\right)  =\phi_{0}\left(  \theta\right)  +\left(
x+1\right)  ^{2}r_{+}\phi_{1}\left(  \theta\right)  /8+\cdots.
\end{equation}
This series expansion naturally imposes $\partial_{x}\phi\left(
x,\theta\right)  =0$ at $x=-1$. With loss of generality, we assume that the
wave function $\phi\left(  x,\theta\right)  $ exhibits definite parity with
respect to the equatorial plane, allowing for restriction of the analysis to
the upper half domain $0\leq\theta\leq\pi/2$. For even and odd parities, the
boundary conditions at $\theta=\pi/2$ are $\partial_{\theta}\phi\left(
x,\theta\right)  =0$ and $\phi\left(  x,\theta\right)  =0$, respectively. The
remaining boundary conditions are
\begin{equation}
\left.  \partial_{\theta}\phi\left(  x,\theta\right)  \right\vert _{\theta
=0}=\phi\left(  1,\theta\right)  =0.
\end{equation}

Using the compactified radial coordinate $x$, the function $\phi\left(
x,\theta\right)  $ is decomposed into a spectral expansion,%
\begin{equation}
\phi\left(  x,\theta\right)  =%
%TCIMACRO{\dsum \limits_{i=0}^{N_{x}-1}}%
%BeginExpansion
{\displaystyle\sum\limits_{i=0}^{N_{x}-1}}
%EndExpansion%
%TCIMACRO{\dsum \limits_{j=0}^{N_{\theta}-1}}%
%BeginExpansion
{\displaystyle\sum\limits_{j=0}^{N_{\theta}-1}}
%EndExpansion
\alpha_{ij}T_{i}\left(  x\right)  \Theta_{j}\left(  \theta\right)  ,
\label{eq:sexpansion}%
\end{equation}
where $N_{x}$ and $N_{\theta}$ denote the resolutions in the radial and
angular coordinates, respectively, $T_{i}\left(  x\right)  $ represents the
Chebyshev polynomial, and $\alpha_{ij}$ are the spectral coefficients. The
angular basis $\Theta_{j}\left(  \theta\right)  $ depends on the parity with
respect to $\theta=\pi/2$. Specifically, we adopt%
\begin{equation}
\Theta_{j}\left(  \theta\right)  =\left\{
\begin{array}
[c]{c}%
\cos\left(  2j\theta\right)  \text{ \ for even parity }\\
\cos\left[  \left(  2j+1\right)  \theta\right]  \text{ for odd parity}%
\end{array}
\right.  \text{.} \label{eq:sexpansion-1-1}%
\end{equation}
This choice automatically satisfies the boundary conditions at $\theta=0$ and
$\pi/2$. To ensure numerical precision and efficiency, we set $\left(
N_{x},N_{\theta}\right)  =\left(  50,5\right)  $ for subsequent numerical
computations of $\phi\left(  x,\theta\right)  $.

To determine the spectral coefficients $\alpha_{ij}$, the spectral expansion
$\left(  \ref{eq:sexpansion}\right)  $ is substituted into the PDE, followed
by discretization at the Gauss-Chebyshev points. This procedure reduces the
PDE for $\phi\left(  x,\theta\right)  $ to a finite system of algebraic
equations involving $\alpha_{ij}$. To circumvent the linear scaling invariance
of Eq. $\left(  \ref{eq:delta phi}\right)  $, a non-trivial solution for
$\alpha_{ij}$ is obtained by imposing $\phi\left(  x,\theta\right)  =1$ at
$\left(  x,\theta\right)  =\left(  -1,0\right)  $. This constraint introduces
an additional algebraic equation for $\alpha_{ij}$ through the spectral
expansion $\left(  \ref{eq:sexpansion}\right)  $. To establish an equal number
of variables and equations, one black hole parameter (e.g., the reduced black
hole charge $q$) is treated as an additional unknown. The resulting algebraic
equations for $\alpha_{ij}$ and $q$ are solved iteratively using the
Newton-Raphson method, with the linear system of equations at each iteration
solved using the built-in LinearSolve command in Mathematica. The
Newton-Raphson algorithm is applied iteratively until successive iterations
converge to within a tolerance of $10^{-10}$.

To delineate the parameter space for KN black holes supporting scalar clouds,
a systematic exploration of scalar cloud solutions is conducted within the
$\left(  \alpha,\chi,q\right)  $ parameter space. The process initiates by
employing the $\phi\left(  x,\theta\right)  $ and $q$ values at the
bifurcation points of RN black holes as seed solutions for the iterative
solver. Subsequently, $\alpha$ or $\chi$ is incrementally adjusted by a
specified step size, with the resulting solution serving as the initial guess
for the next iteration. This iterative procedure continues until the
identification of additional solutions becomes computationally infeasible,
indicating the boundary of the parameter space. Near the parameter space
boundary, the step size is adaptively refined to enhance accuracy. Throughout
the iterative process, the residual of the spectral approximation and the
number of nodes are monitored to ensure solution accuracy. A residual
tolerance of $10^{-7}$ is generally maintained.

\section{Results}

\label{sec:Numerical Results}

This section presents numerical results regarding the parameter space of KN
black holes that can admit scalar clouds for the fundamental and first two
excited modes. Representative examples of scalar cloud wave functions are also provided.

\subsection{Fundamental Mode}

\begin{figure}[ptb]
\begin{centering}
\includegraphics[scale=0.5]{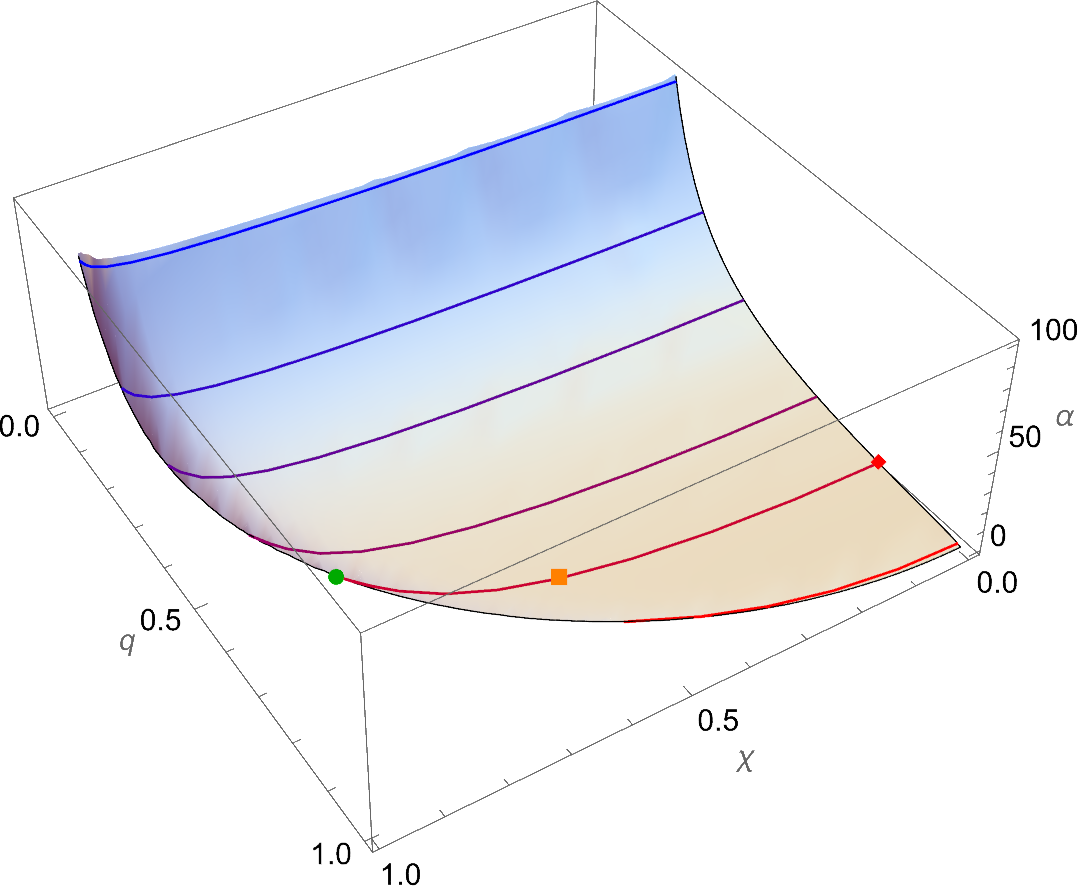} \includegraphics[scale=0.7]{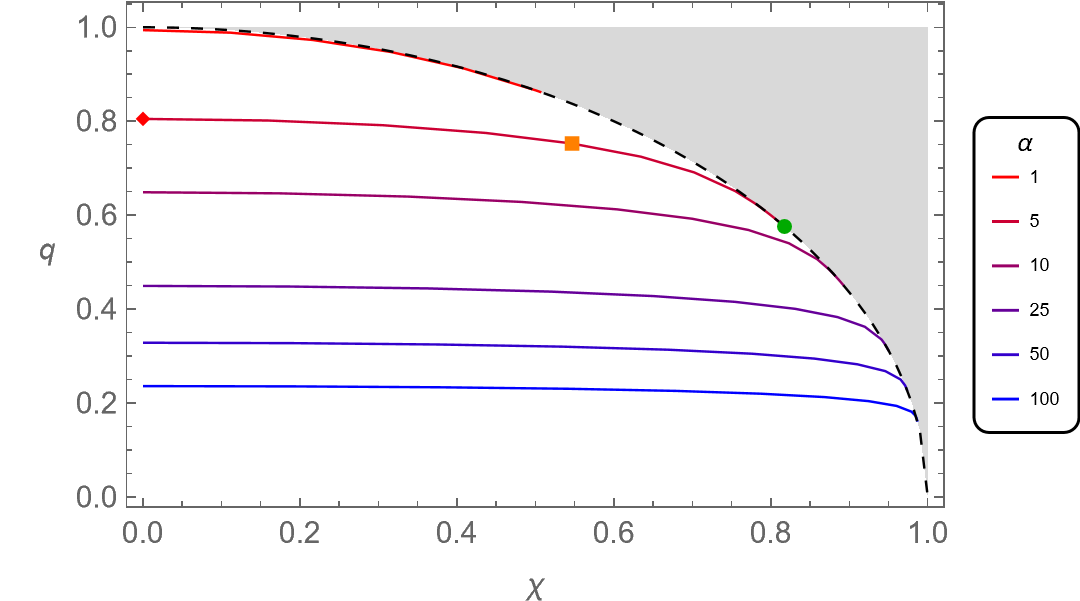}
\includegraphics[scale=0.42]{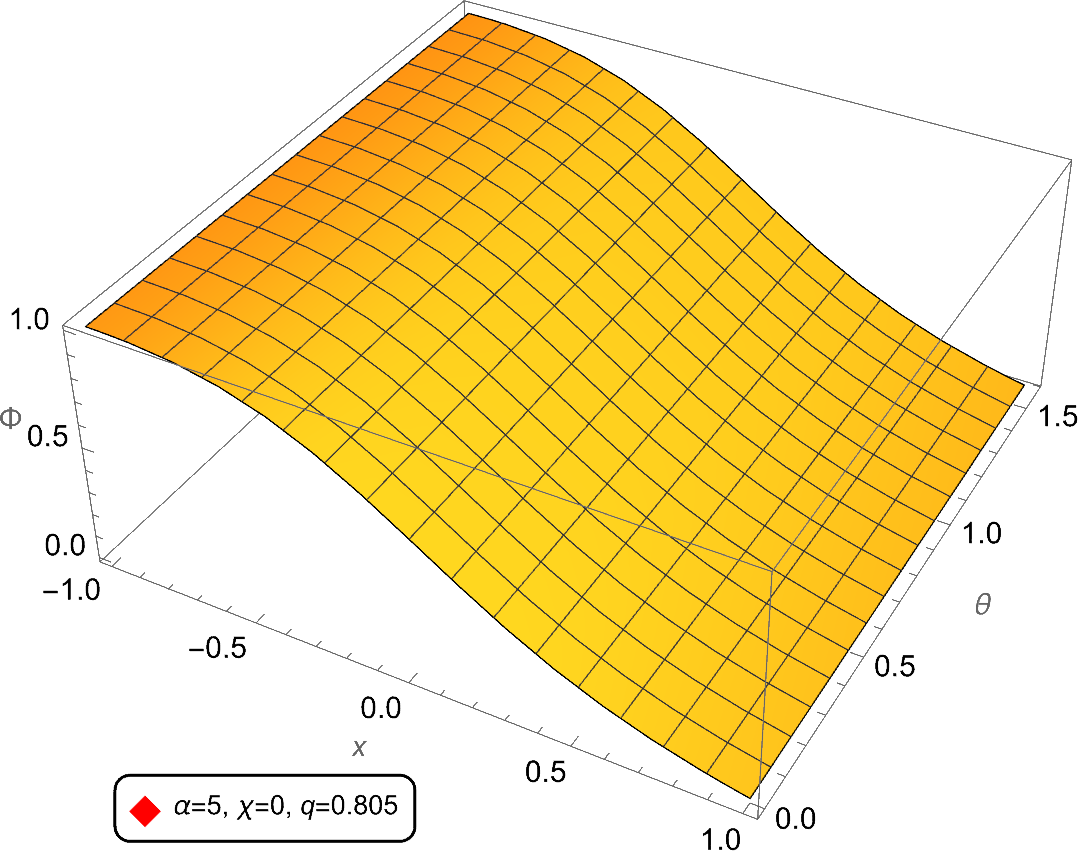} \includegraphics[scale=0.42]{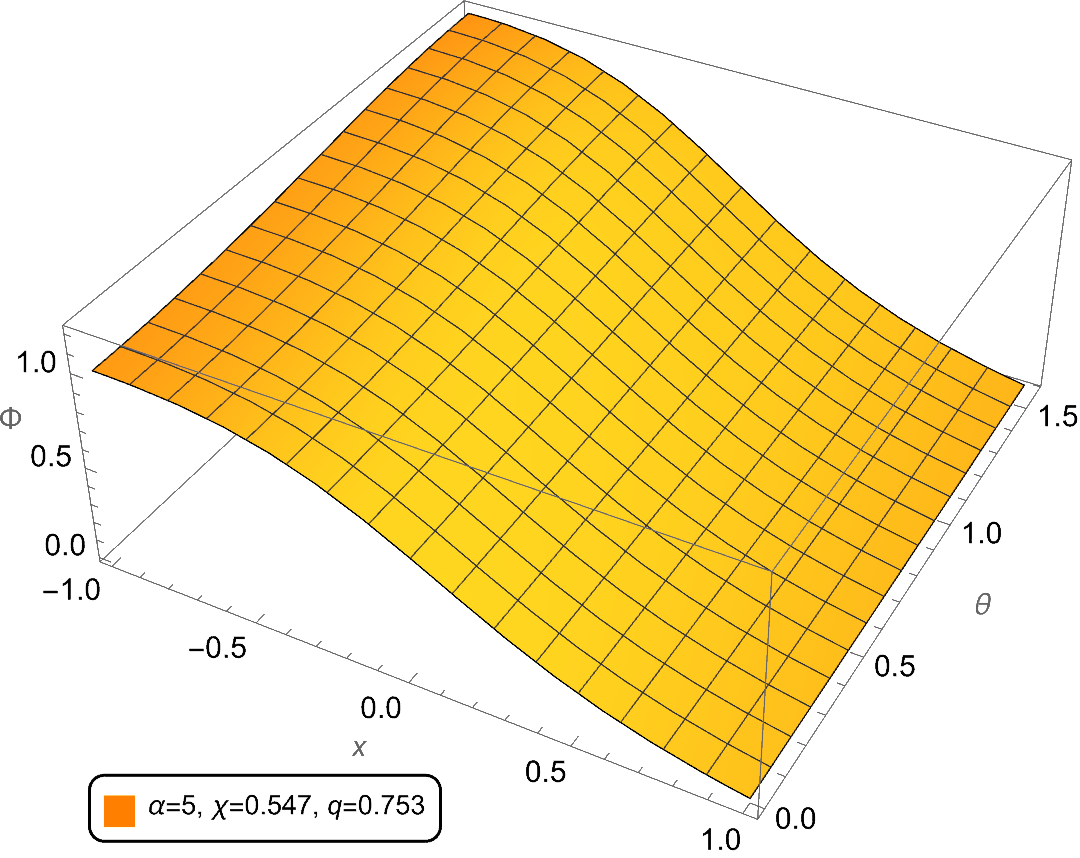}
\includegraphics[scale=0.42]{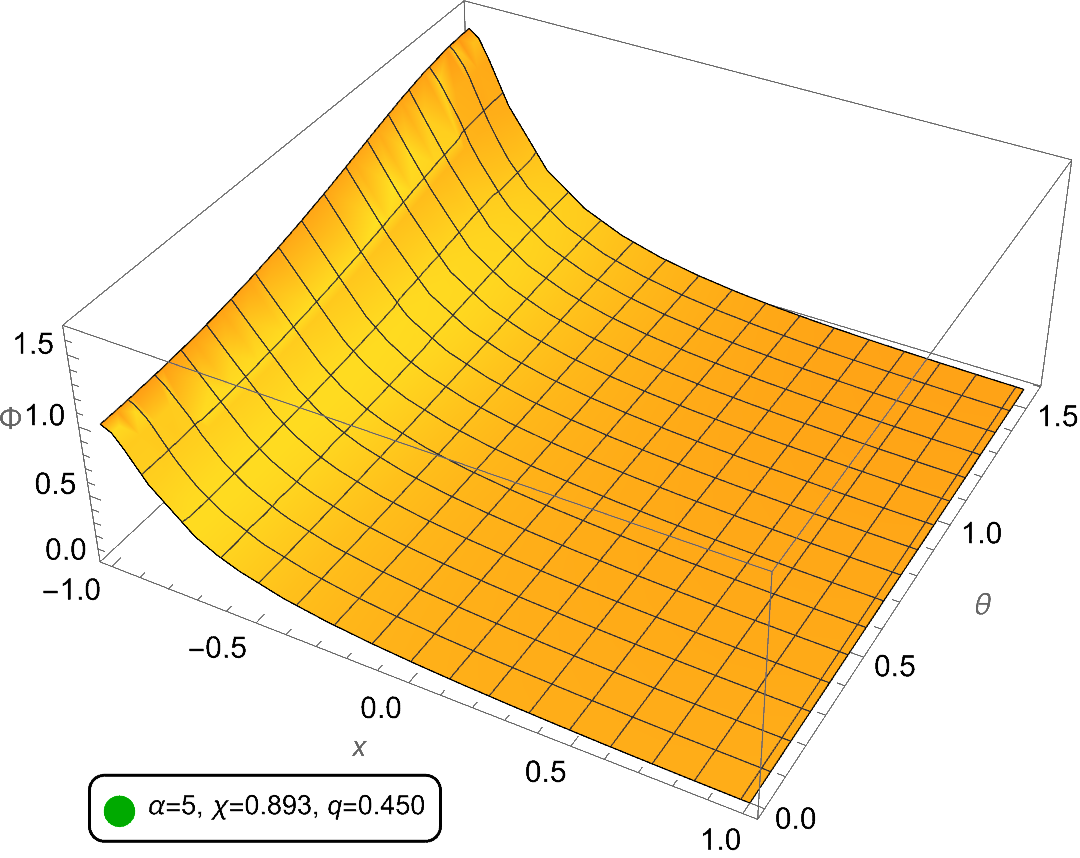}
\par\end{centering}
\caption{\textbf{Upper-Left Panel:} Existence domain of fundamental scalar
clouds with $\left(  n,l\right)  =\left(  0,0\right)  $ in the $\left(
\alpha,\chi,q\right)  $ parameter space. KN black holes residing on the
colored surface admit the scalar clouds. Existence lines for $\alpha=1$, $5$,
$10$, $25$, $50$ and $100$ are shown, alongside three representative examples
on the $\alpha=5$ line. \textbf{Upper-Right Panel:} Existence line for
$\alpha=1$, $5$, $10$, $25$, $50$ and $100$ in the $\left(  \chi,q\right)  $
space. These lines approach and terminate at the extremal KN black hole line
(black dashed), beyond which no KN black holes exist (gray region).
\textbf{Lower Row:} Wave function $\phi\left(  x,\theta\right)  $ in three
representative $\alpha=5$ KN black holes. The non-rotating black hole exhibits
spherically symmetric scalar clouds. Rotating black holes demonstrate a
concentration of scalar clouds near the event horizon and equatorial plane.}%
\label{fig:Fwaveform}%
\end{figure}

We begin by examining the fundamental mode of scalar clouds with $\left(
n,l\right)  =\left(  0,0\right)  $, characterized by nodeless wave functions.
Nonlinear realizations of these $\left(  n,l\right)  =\left(  0,0\right)  $
clouds correspond to scalarized KN black holes in the fundamental state,
previously investigated in \cite{Guo:2023mda}. The upper-left panel of Fig.
\ref{fig:Fwaveform} depicts the existence domain for the fundamental clouds
within the $\left(  \alpha,\chi,q\right)  $ parameter space, where KN black
holes supporting such clouds reside on the colored surface. Additionally,
existence lines for various fixed $\alpha$ values are presented, accompanied
by three representative scalar cloud profiles with $\alpha=5$. The upper-right
panel of Fig. \ref{fig:Fwaveform} illustrates the same existence lines in the
$\left(  \chi,q\right)  $ plane. The black dashed line represents the extremal
limit, corresponding to KN black holes satisfying $q^{2}+\chi^{2}=1$. The gray
region above this extremal line is inaccessible to KN black holes, imposing an
upper bound on the black hole charge for a given $\chi$. As $\chi$ increases,
this upper bound decreases from unity to zero. It is noteworthy that the
effective mass squared $\mu_{\text{eff}}^{2}$ becomes increasingly negative
for a given $\chi$ as $\alpha$ or $q$ grows (cf. Eq. $\left(  \ref{eq:negmass}%
\right)  $), signifying an amplification of tachyonic instabilities with
larger $\alpha$ or $q$ values.

Four key characteristics are observed regarding the existence lines:

\begin{itemize}
\item \textbf{Shift Toward Smaller }$q$\textbf{ with Increasing }$\alpha
$\textbf{:} As $\alpha$ increases, the existence lines shift towards smaller
$q$ values. This is attributed to the enhancement of tachyonic instabilities
for larger $\alpha$, thereby allowing a lower $q$ to induce scalar cloud formation.

\item \textbf{Approach to Extremal Line with Increasing }$\chi$\textbf{:} With
increasing $\chi$, each existence line approaches the extremal line and
ultimately terminates at this boundary. Consequently, the charge of the
existence lines converges to its upper limit as black hole spin accelerates.
This implies that black hole spin suppresses scalar cloud formation, and
scalar clouds cease to exist beyond a critical spin, denoted by $\chi
_{\text{ext}}\left(  \alpha\right)  $.

\item \textbf{Threshold Line for Scalarized KN Black Holes: }For a given
$\alpha$, the corresponding existence line serves as a threshold for
scalarized KN black holes. KN black holes below this line exhibit insufficient
tachyonic instabilities to support scalar clouds, let alone scalarized black
holes, due to their low $q$ values. Conversely, KN black holes above the
existence line possess excessively strong tachyonic instabilities,
necessitating nonlinear effects to suppress these instabilities and give rise
to scalarized KN black holes.

\item \textbf{Decrease of Existence Line with Increasing }$\chi$\textbf{:} For
a fixed $\alpha$, the existence line exhibits a slight decrease as $\chi$
increases from zero towards the extremal line. At $\chi=0$, the $q$ value of
the existence line, denoted by $q_{\text{RN}}\left(  \alpha\right)  $,
corresponds to the bifurcation point of the RN black hole with the given
$\alpha$. At the extremal line termination point, the existence line yields a
threshold charge, denoted by $q_{\text{ext}}\left(  \alpha\right)  $. Notably,
$q_{\text{RN}}\left(  \alpha\right)  $ consistently exceeds $q_{\text{ext}%
}\left(  \alpha\right)  $ for fundamental scalar clouds. Tachyonic
instabilities of KN black holes with $q<$ $q_{\text{ext}}\left(
\alpha\right)  $ and $q>q_{\text{RN}}\left(  \alpha\right)  $ are too weak and
strong, respectively, to allow the existence of scalar clouds.
\end{itemize}

The lower row of Fig. \ref{fig:Fwaveform} presents the wave function
$\phi\left(  x,\theta\right)  $ of fundamental scalar clouds for three KN
black holes situated on the $\alpha=5$ existence line. The wave function
$\phi\left(  x,\theta\right)  $ in the lower-left panel corresponds to the RN
black hole bifurcation point and exhibits $\theta$-independence due to the
spherical symmetry of RN black holes. The lower-middle and lower-right panels
demonstrate a notable elevation of the wave function $\phi\left(
x,\theta\right)  $ on the equatorial plane as black hole spin increases.
Furthermore, the wave function at the extremal limit displays a sharper peak
at the event horizon compared to the cases with zero or moderate spin. These
observations collectively indicate that black hole spin induces a
concentration of scalar clouds towards both the horizon and the equatorial plane.

\begin{figure}[ptb]
\begin{centering}
\includegraphics[scale=0.37]{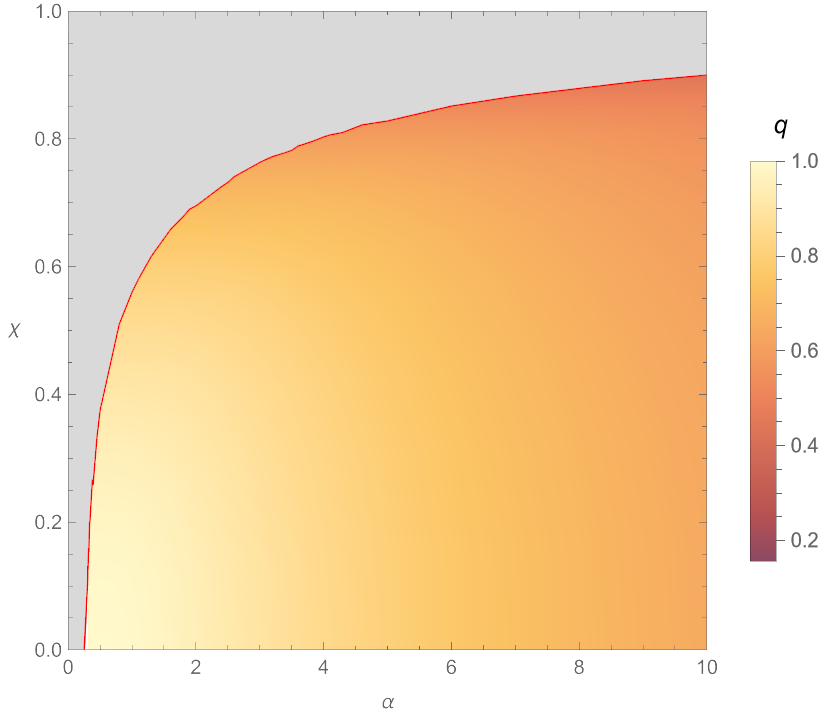} \includegraphics[scale=0.37]{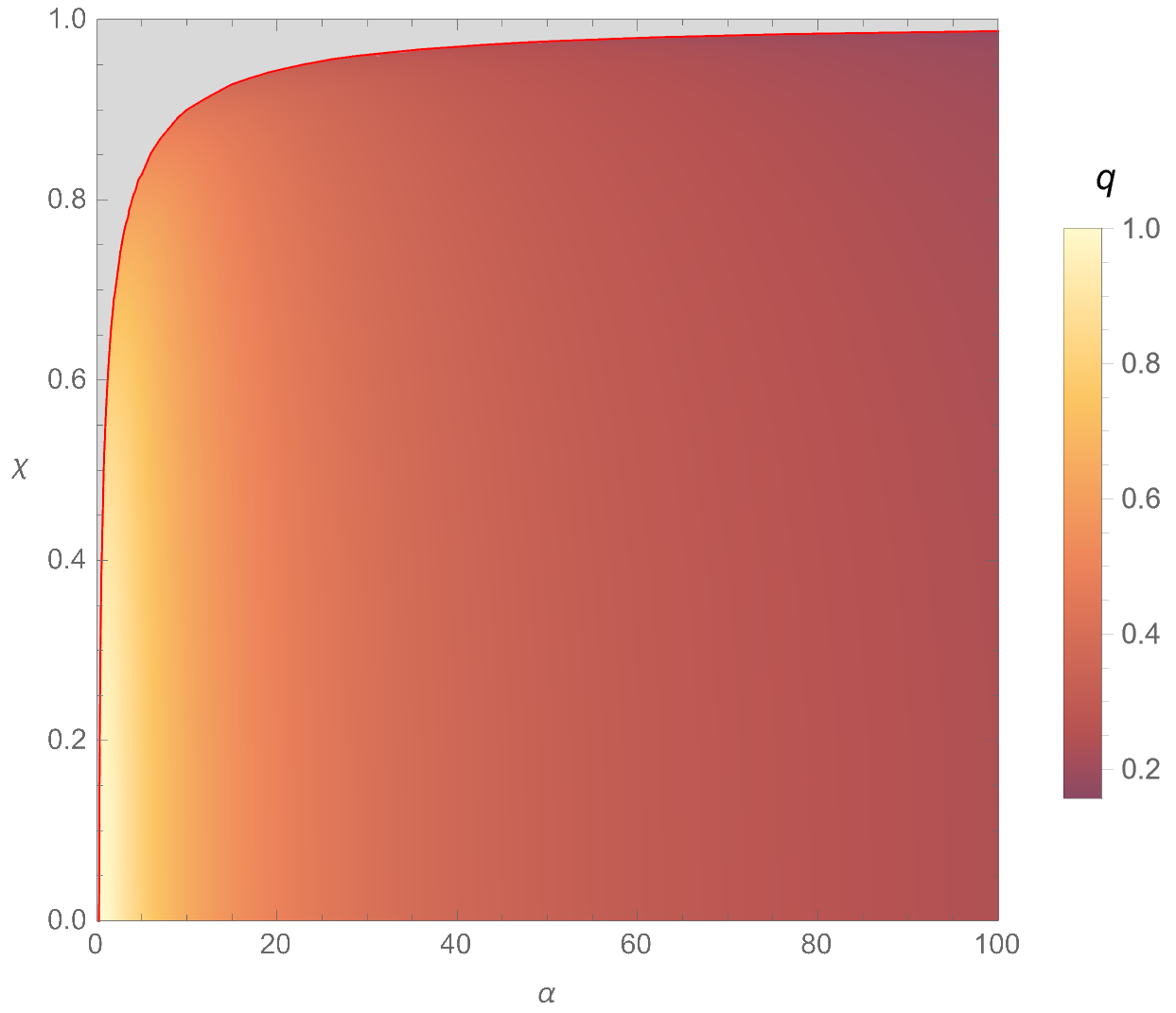}
\includegraphics[scale=0.37]{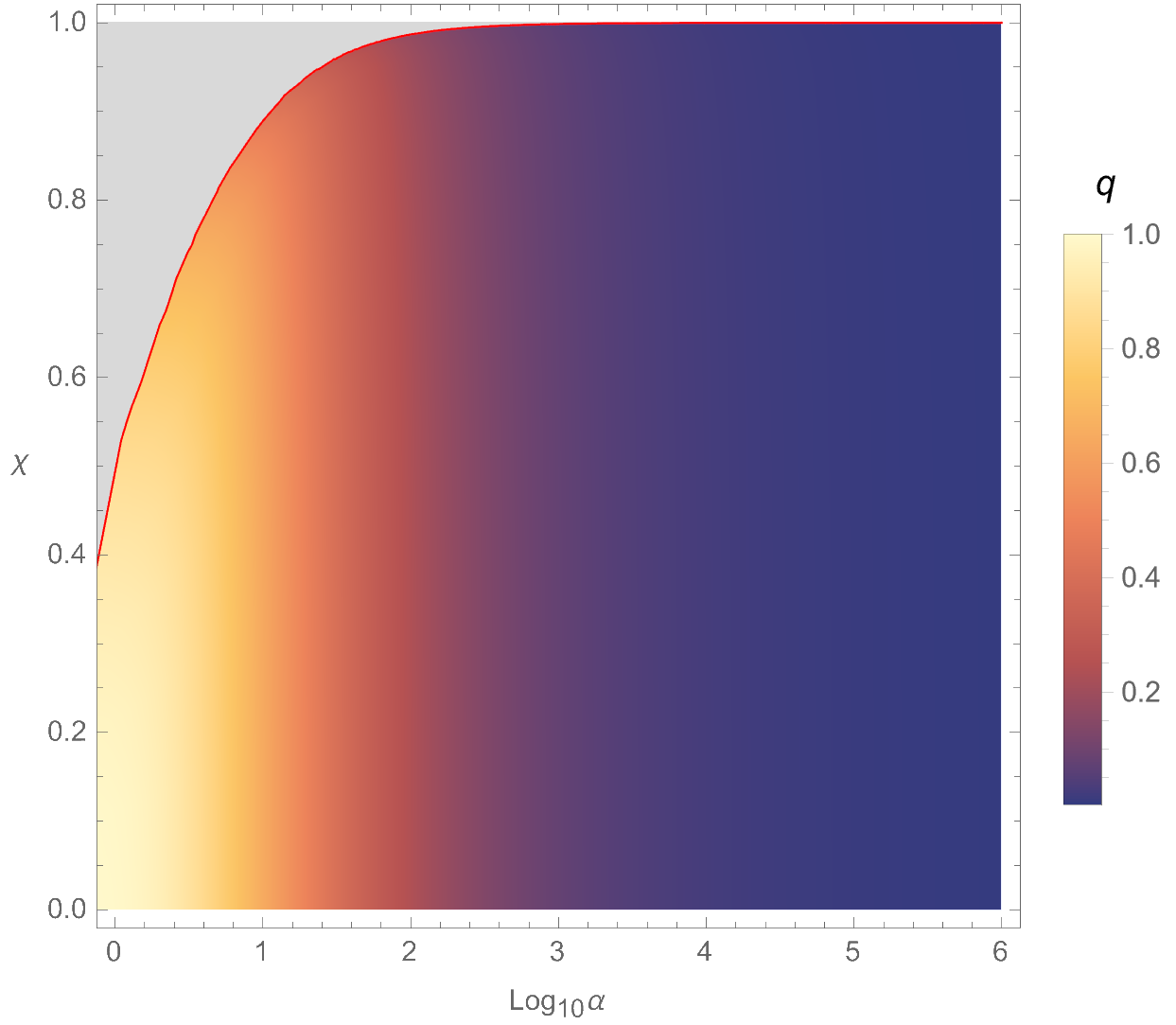} \includegraphics[scale=0.37]{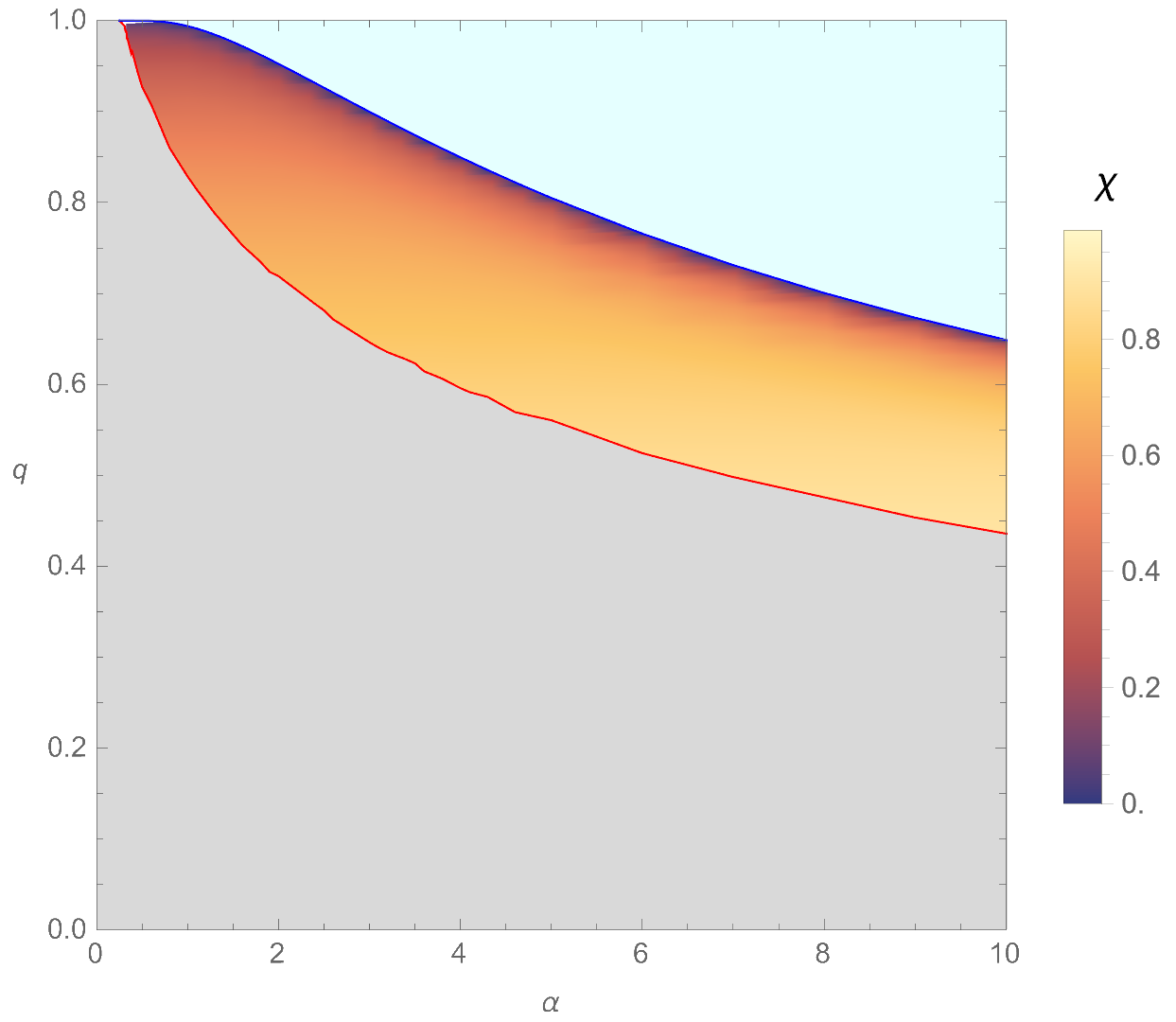}
\includegraphics[scale=0.37]{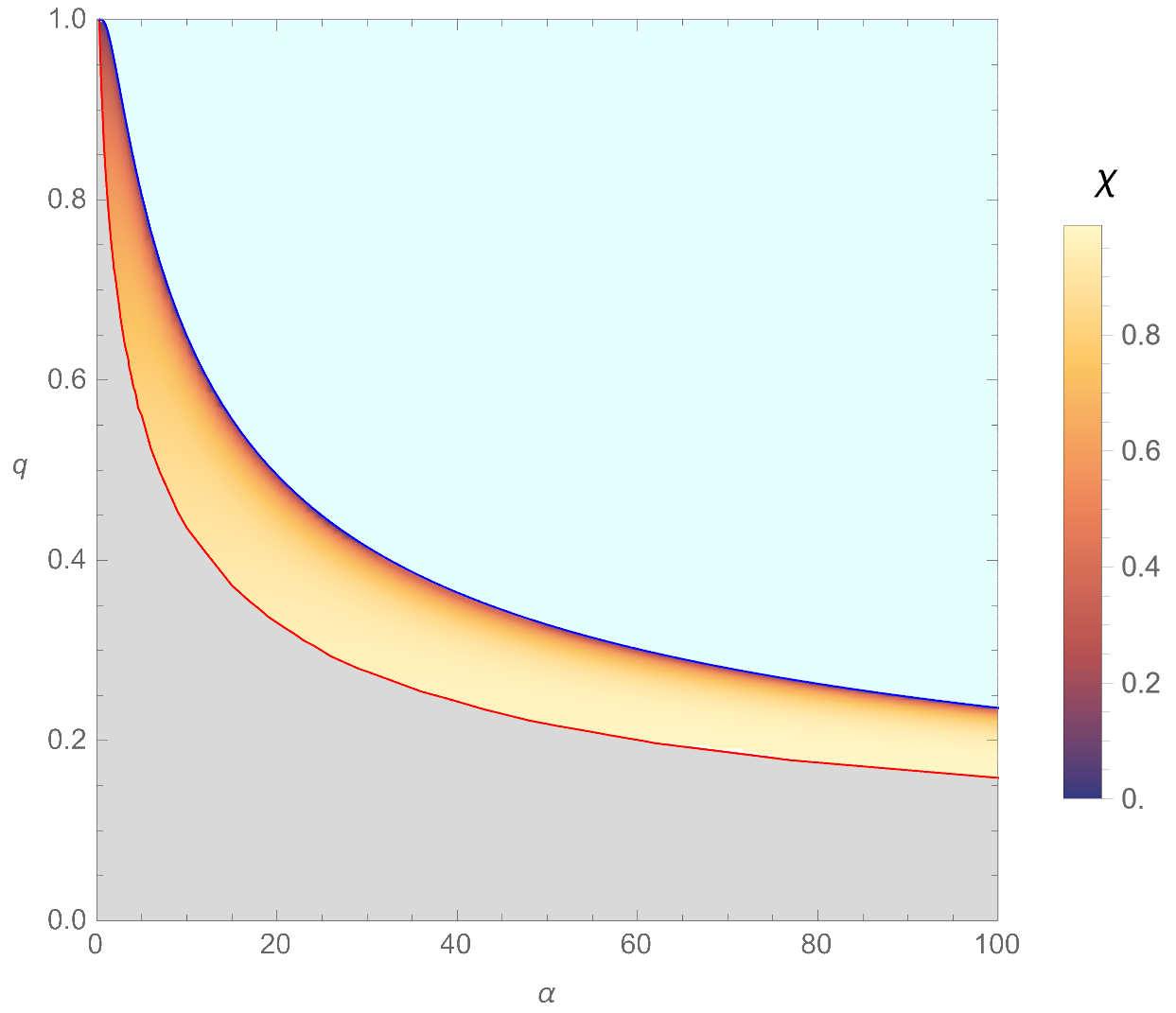} \includegraphics[scale=0.37]{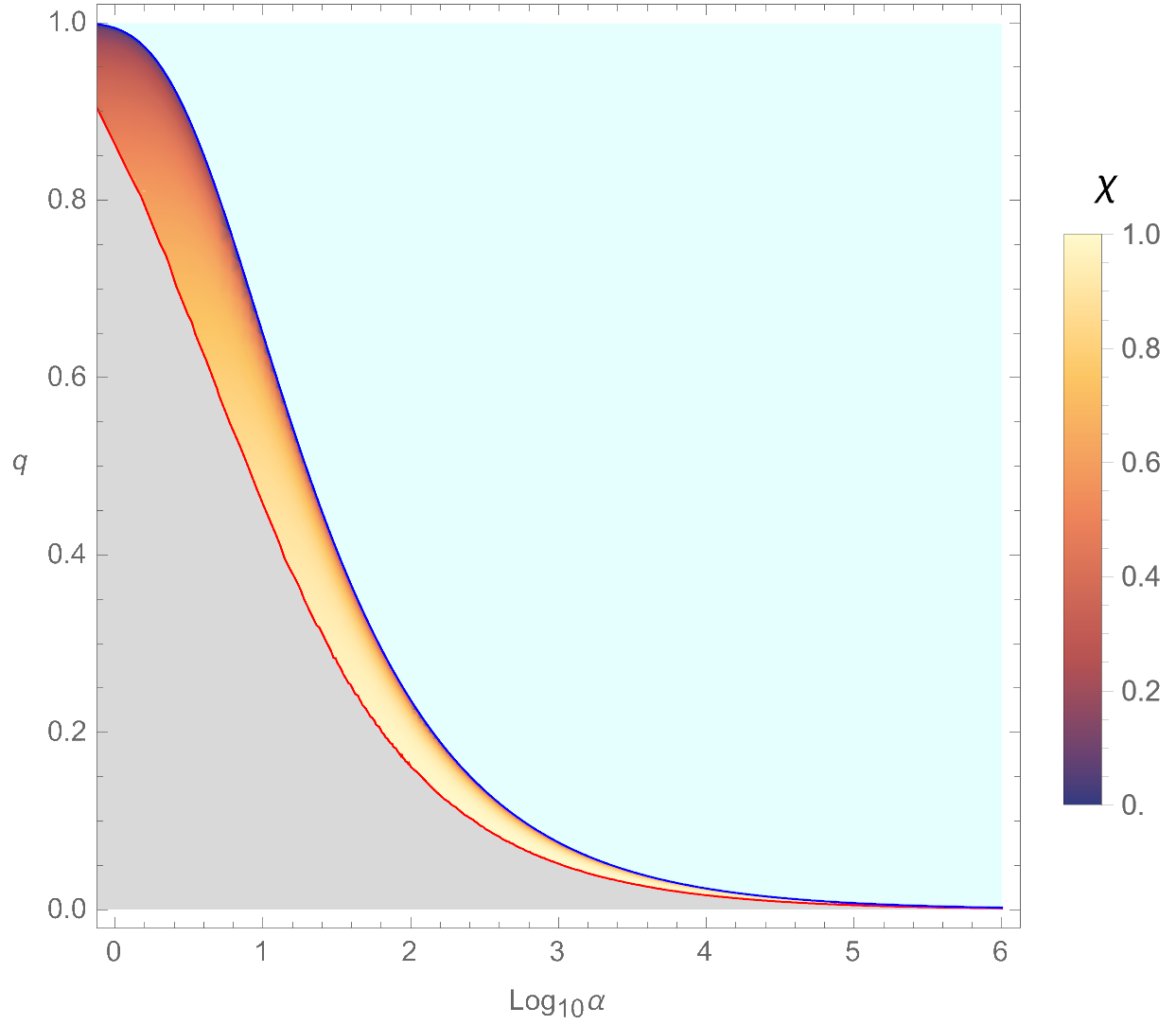}
\par\end{centering}
\caption{\textbf{Upper Row}: Existence domain of fundamental scalar clouds in
the $\left(  \alpha,\chi\right)  $ plane for various $\alpha$ ranges. Density
plot colors represent $q$ values. The domain is bounded by the red line,
composed of extremal KN black holes. Scalar clouds cannot form in the gray
region due to insufficient tachyonic instabilities. \textbf{Lower Row:
}Existence domain in the $\left(  \alpha,q\right)  $ plane for various
$\alpha$ ranges. Density plot colors correspond to $\chi$ values. The domain
is bounded by the blue and red lines, which represent RN and extremal KN black
holes, respectively. Scalar clouds cannot form in gray regions due to
insufficient tachyonic instabilities, while those in cyan regions experience
excessively strong instabilities to sustain a stationary state.}%
\label{fig:Fpara}%
\end{figure}

The upper and lower rows of Fig. \ref{fig:Fpara} present density plots of the
fundamental cloud existence domain in the $\left(  \alpha,\chi\right)  $ and
$\left(  \alpha,q\right)  $ spaces, respectively. Color variations within the
density plots indicate corresponding $q$ or $\chi$ values. The existence
domain in the $\left(  \alpha,\chi\right)  $ space is bounded by the upper
limit $\chi_{\text{ext}}\left(  \alpha\right)  $, represented by red lines.
Regions above these red lines, denoted as gray, preclude scalar cloud
formation as black hole charges must exceed their extremal values to induce
sufficient tachyonic instabilities. As anticipated, smaller $q$ values are
required to sustain scalar clouds for stronger coupling constants $\alpha$.
Furthermore, the existence domain in the $\left(  \alpha,q\right)  $ space is
confined by the upper bound $q_{\text{RN}}\left(  \alpha\right)  $, depicted
as blue lines, and the lower bound $q_{\text{ext}}\left(  \alpha\right)  $,
represented by red lines. Similarly, KN black holes within gray regions below
the red lines cannot support scalar clouds due to insufficient tachyonic
instabilities. Conversely, cyan regions above the blue line exhibit overly
strong tachyonic instabilities to accommodate scalar clouds, leading instead
to the emergence of scalarized KN black holes. Moreover, as the coupling
constant $\alpha$ increases, the fundamental scalar cloud existence region
contracts, with both $q_{\text{RN}}\left(  \alpha\right)  $ and $q_{\text{ext}%
}\left(  \alpha\right)  $ approaching zero.

\subsection{Excited Modes}

The preceding subsection explored the fundamental mode of scalar clouds, whose
existence lines mark the onset of fundamental scalarized KN black holes.
Analogously, the presence of excited scalar clouds indicates the appearance of
scalarized KN black holes in excited states, which remain unexplored.
Investigating the existence domain and wave function of excited scalar clouds
can illuminate properties of these scalarized black holes. This subsection
examines two excited states, specifically those with $\left(  n,l\right)
=\left(  0,1\right)  $ and $\left(  1,0\right)  $.

\begin{figure}[ptb]
\begin{centering}
\includegraphics[scale=0.5]{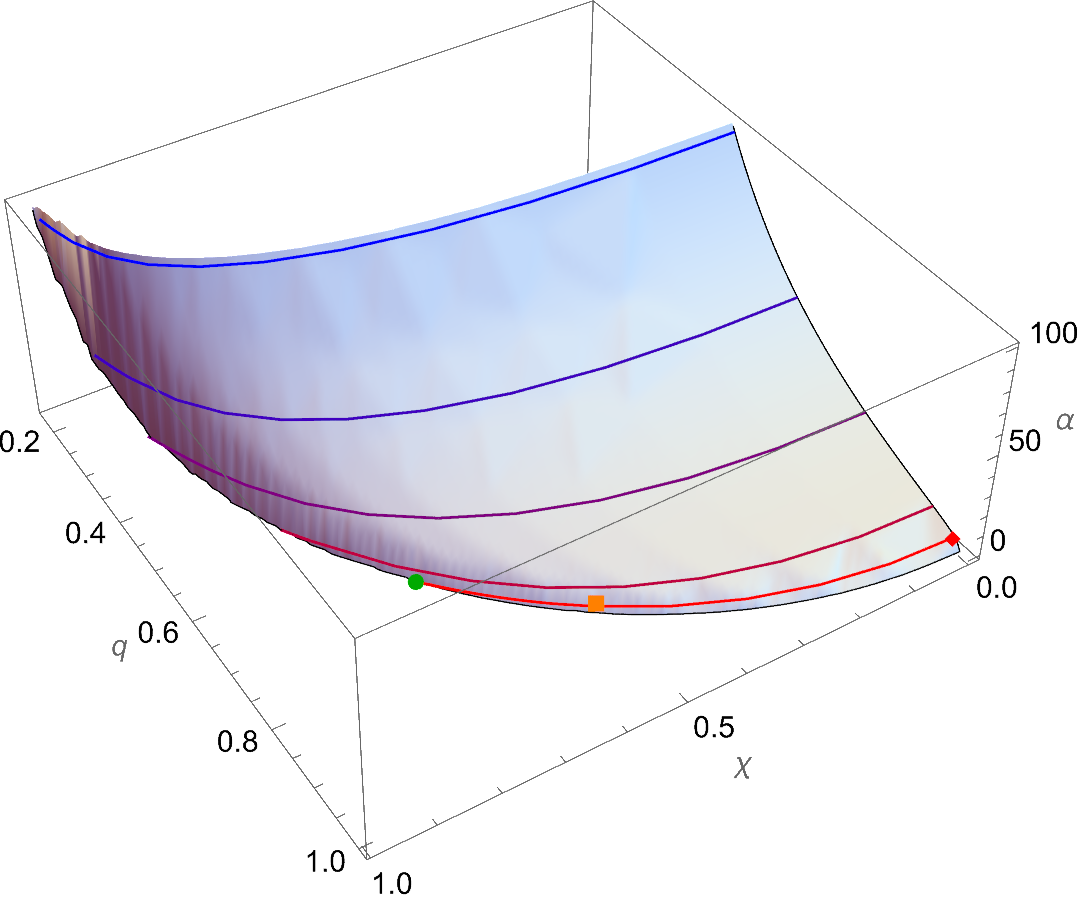} \includegraphics[scale=0.7]{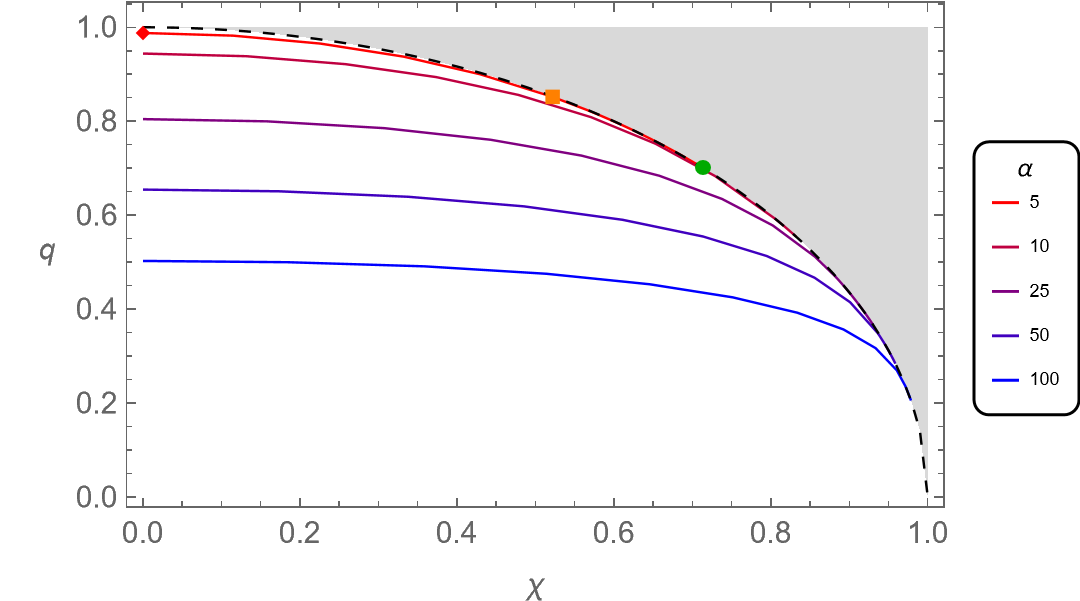}
\includegraphics[scale=0.42]{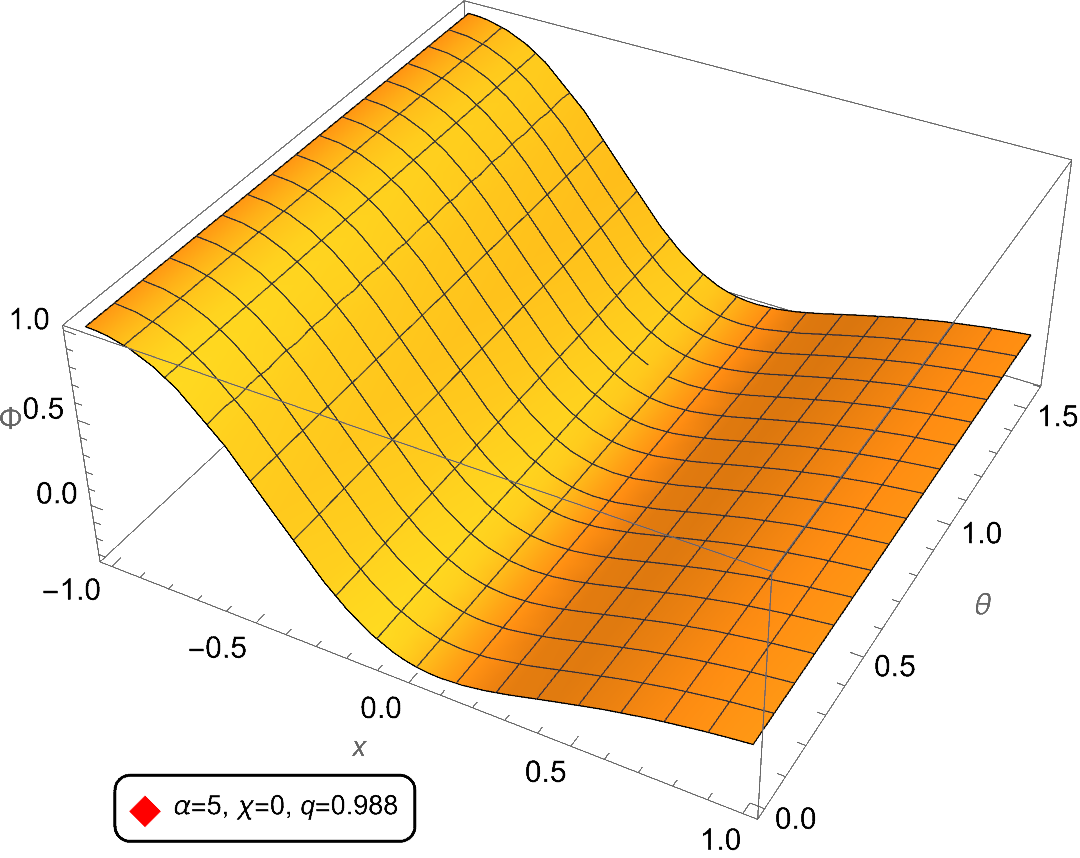} \includegraphics[scale=0.42]{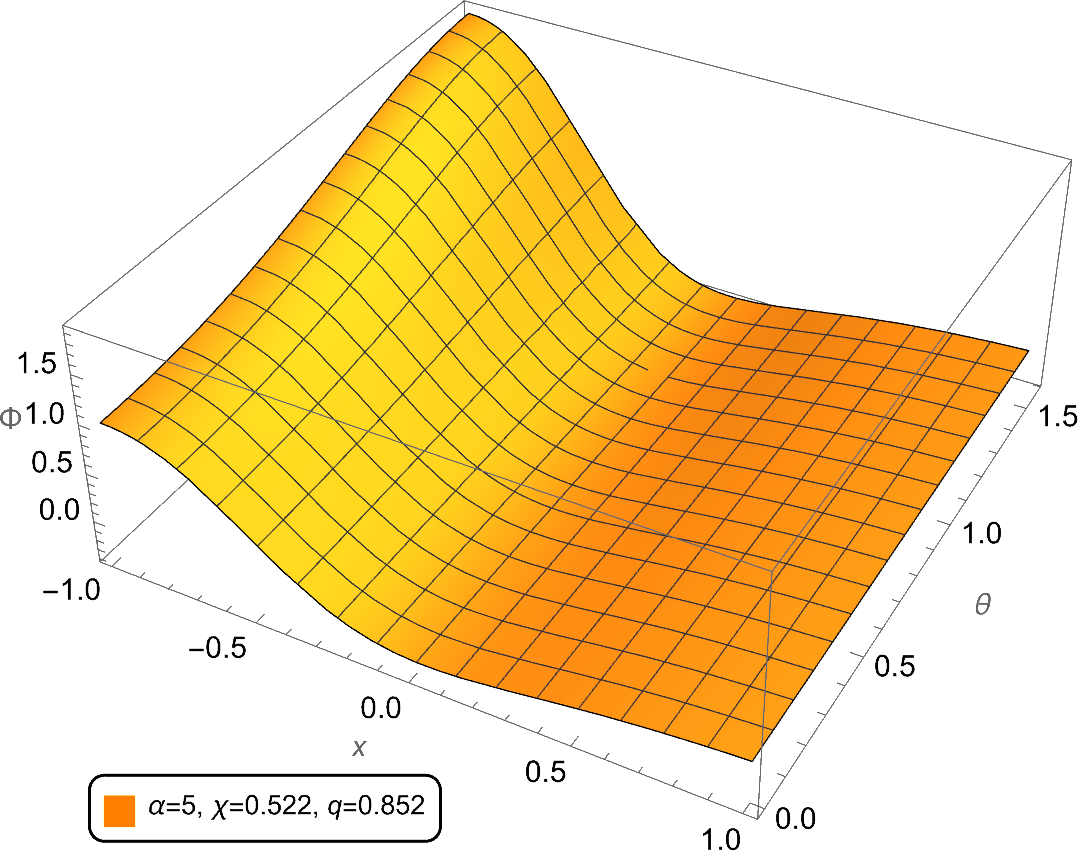}
\includegraphics[scale=0.42]{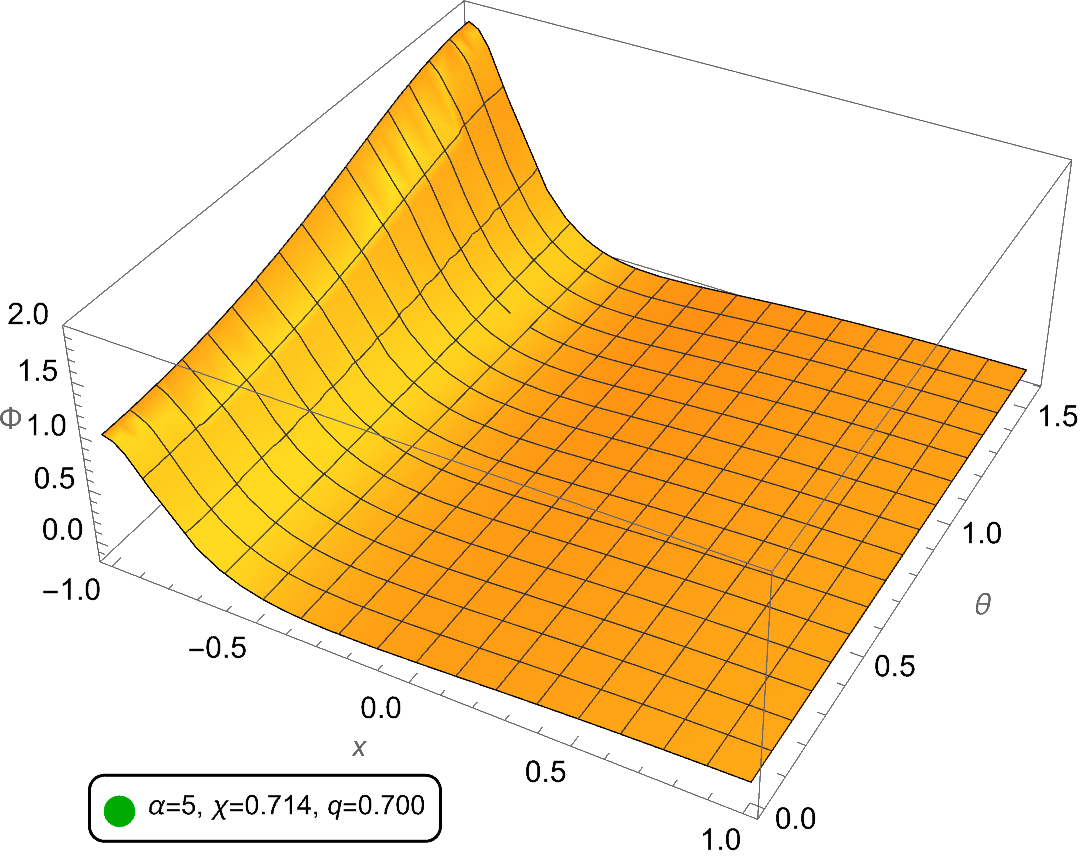}
\par\end{centering}
\caption{Existence domain and wave function of excited scalar clouds with
$\left(  n,l\right)  =\left(  1,0\right)  $. \textbf{Upper-Left Panel}:
Existence domain of excited clouds in the $\left(  \alpha,\chi,q\right)  $
space. \textbf{Upper-Right Panel}: Existence line for $\alpha=$ $5$, $10$,
$25$, $50$ and $100$ in the $\left(  \chi,q\right)  $ space, exhibiting
similarities to the fundamental mode. \textbf{Lower Row:} Wave function of
three representative excited clouds with $\alpha=5$. The radial node results
in a $\theta$-directional valley. Black hole rotation concentrates scalar
clouds near the equatorial plane.}%
\label{fig:Exwaveform10}%
\end{figure}

Fig. \ref{fig:Exwaveform10} illustrates the existence domain and wave function
of scalar clouds with $\left(  n,l\right)  =\left(  1,0\right)  $. The
existence surface in the $\left(  \alpha,\chi,q\right)  $ space and existence
lines in the $\left(  \chi,q\right)  $ space exhibit similarities to the
fundamental mode. However, for a given $\alpha$, the excited mode's existence
line lies above that of the fundamental mode, indicating a higher charge
requirement (and stronger tachyonic instabilities) for excited scalar cloud
formation. The lower row depicts wave functions for three excited clouds with
$\alpha=5$, revealing a valley along the $\theta$ direction due to the
presence of a radial node. As black hole spin increases, this valley
approaches the event horizon. Similar to fundamental clouds, initially
spherically symmetric excited clouds exhibit increased concentration near the
equatorial plane with growing spin.

\begin{figure}[ptb]
\begin{centering}
\includegraphics[scale=0.5]{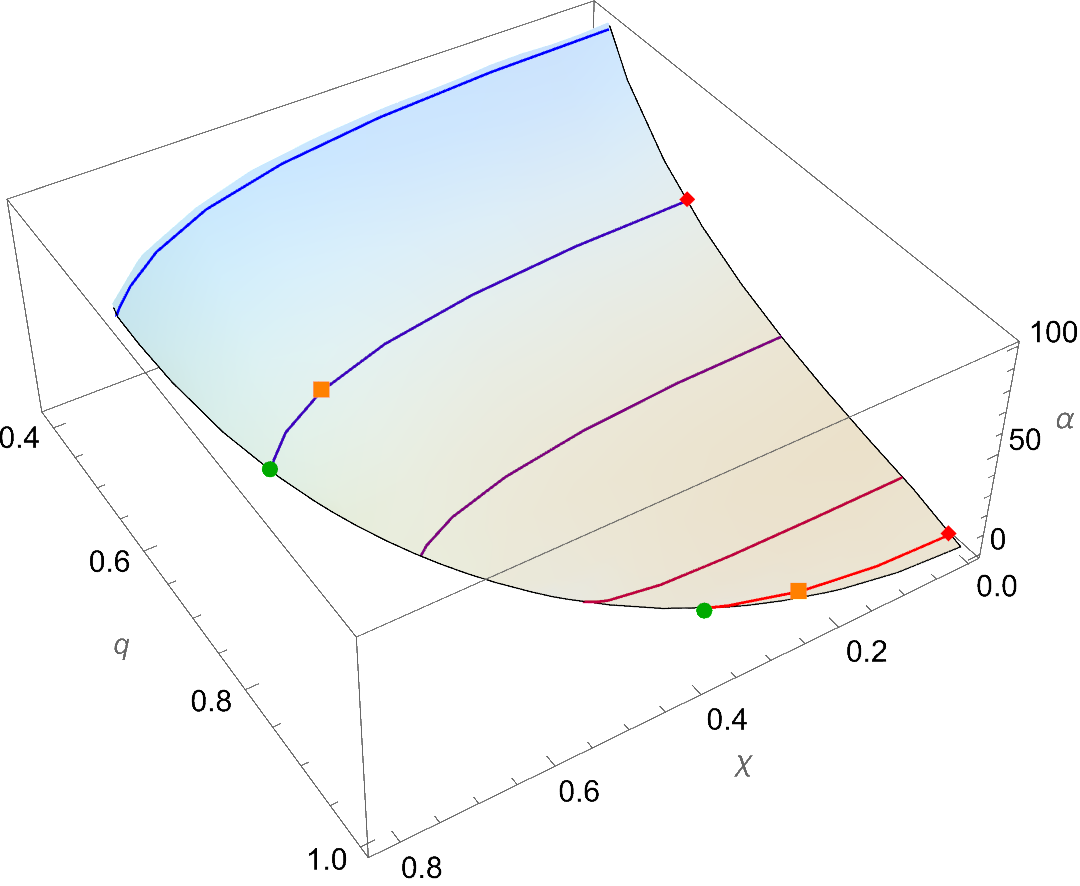} \includegraphics[scale=0.7]{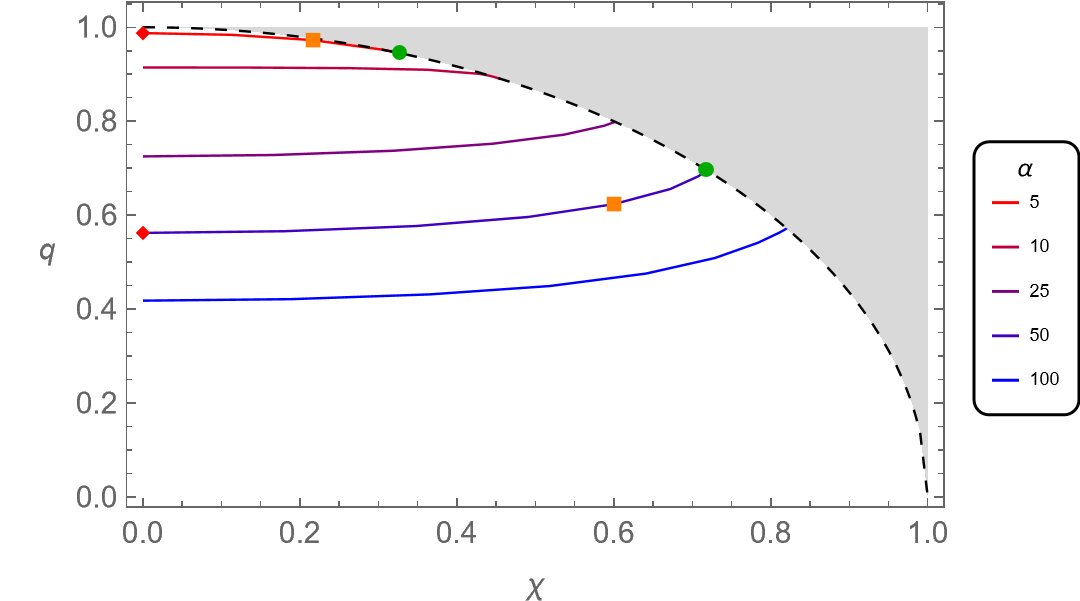}
\includegraphics[scale=0.42]{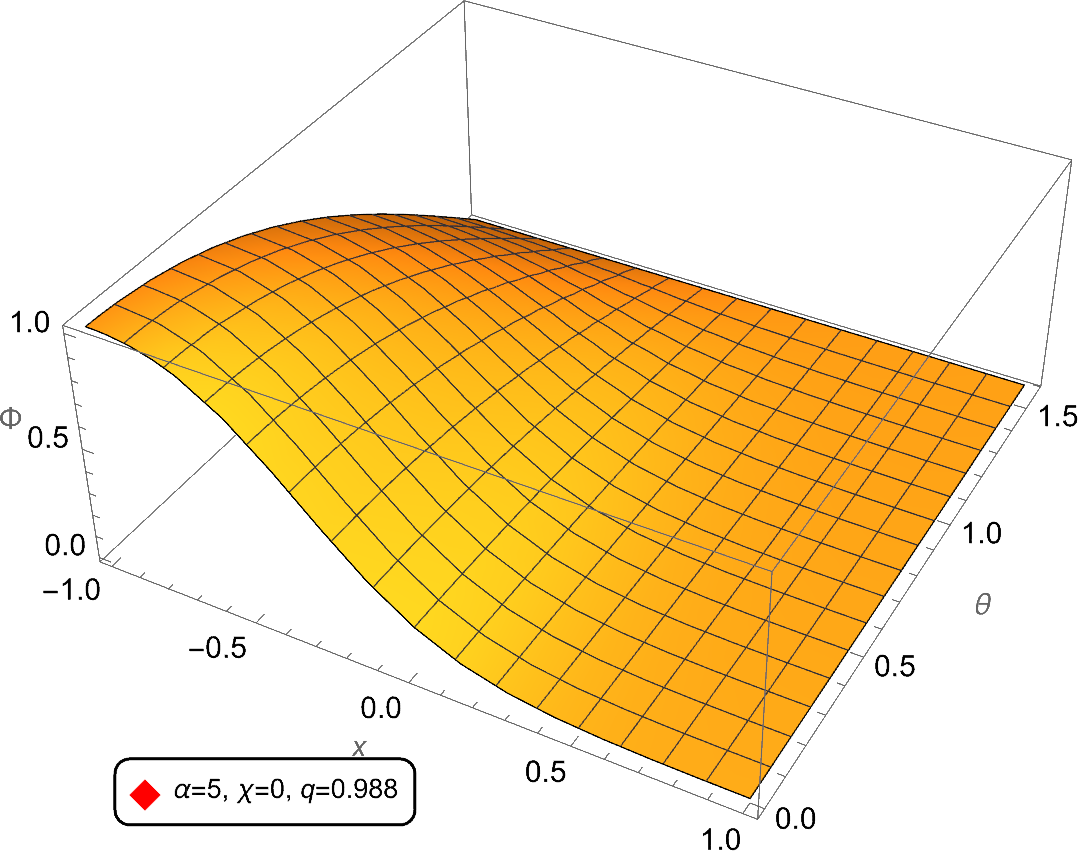} \includegraphics[scale=0.42]{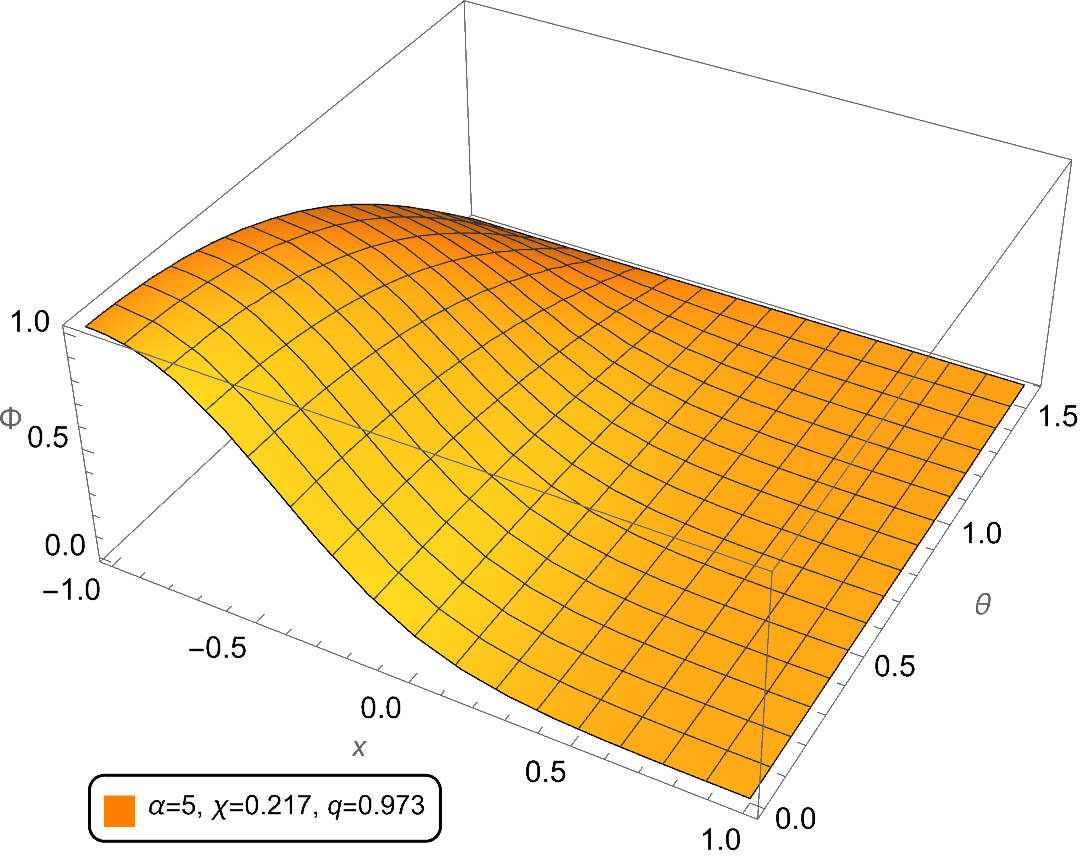}
\includegraphics[scale=0.42]{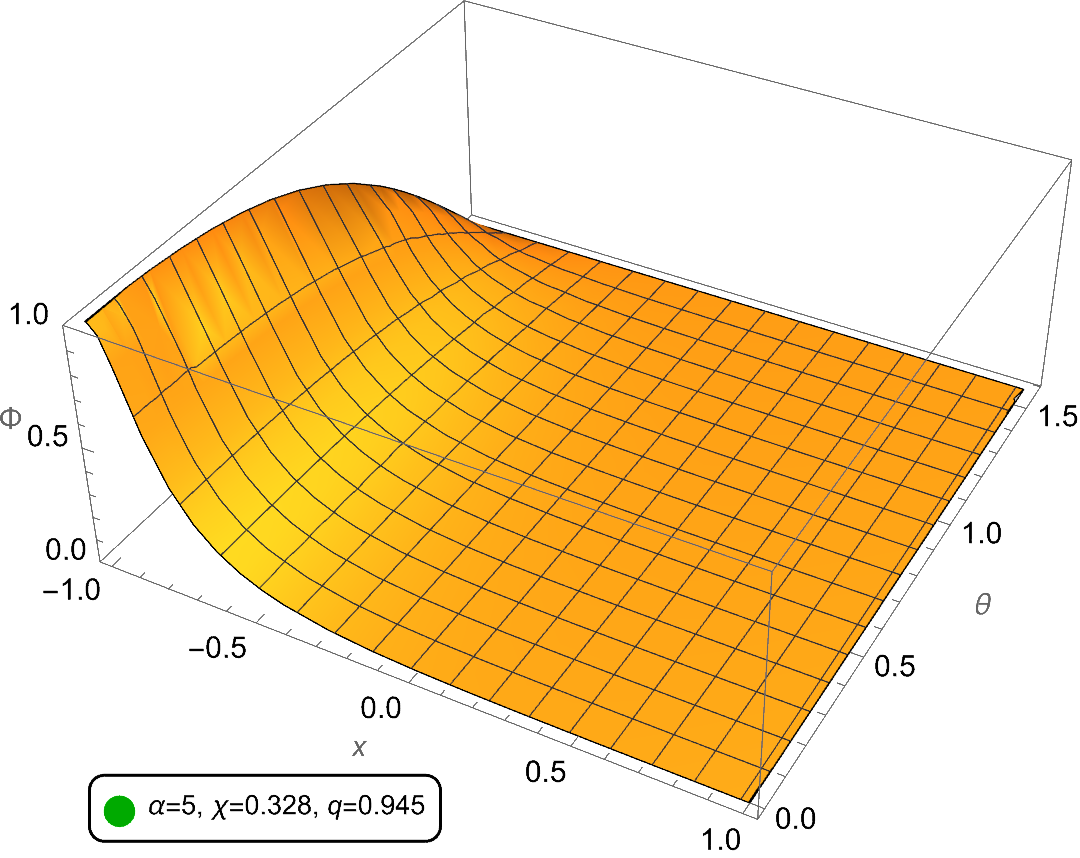} \includegraphics[scale=0.42]{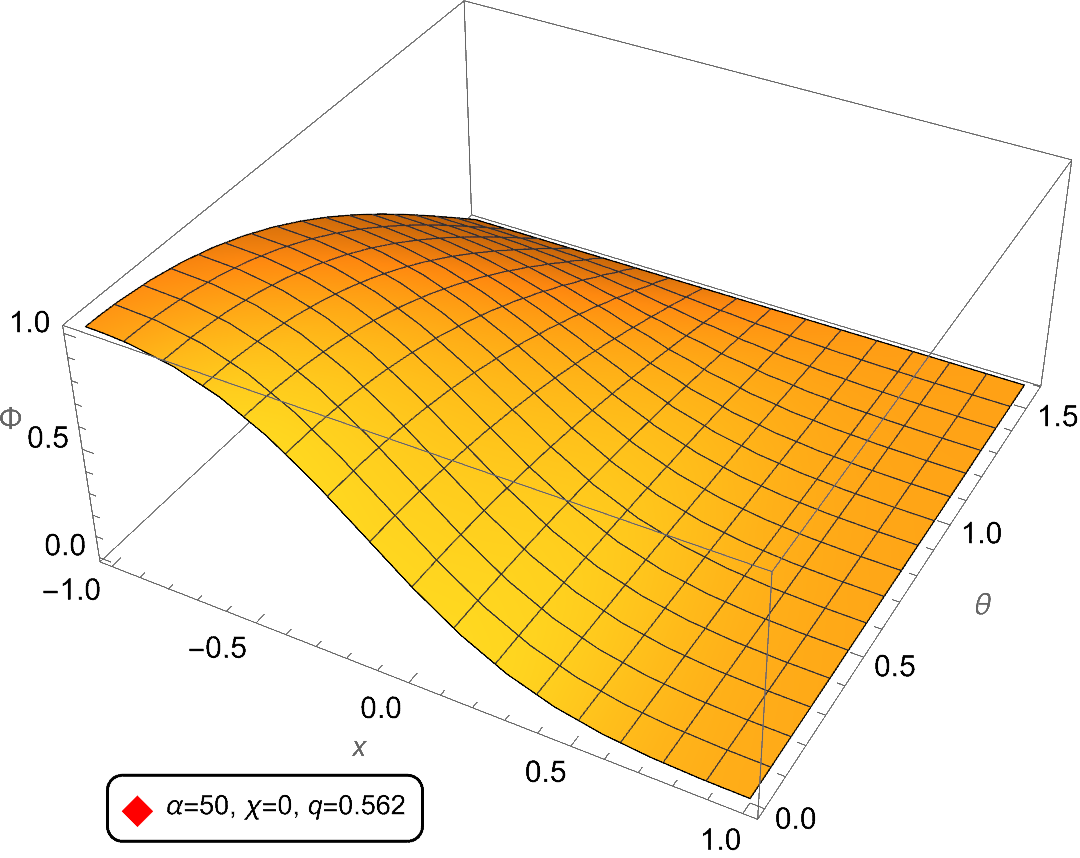}
\includegraphics[scale=0.42]{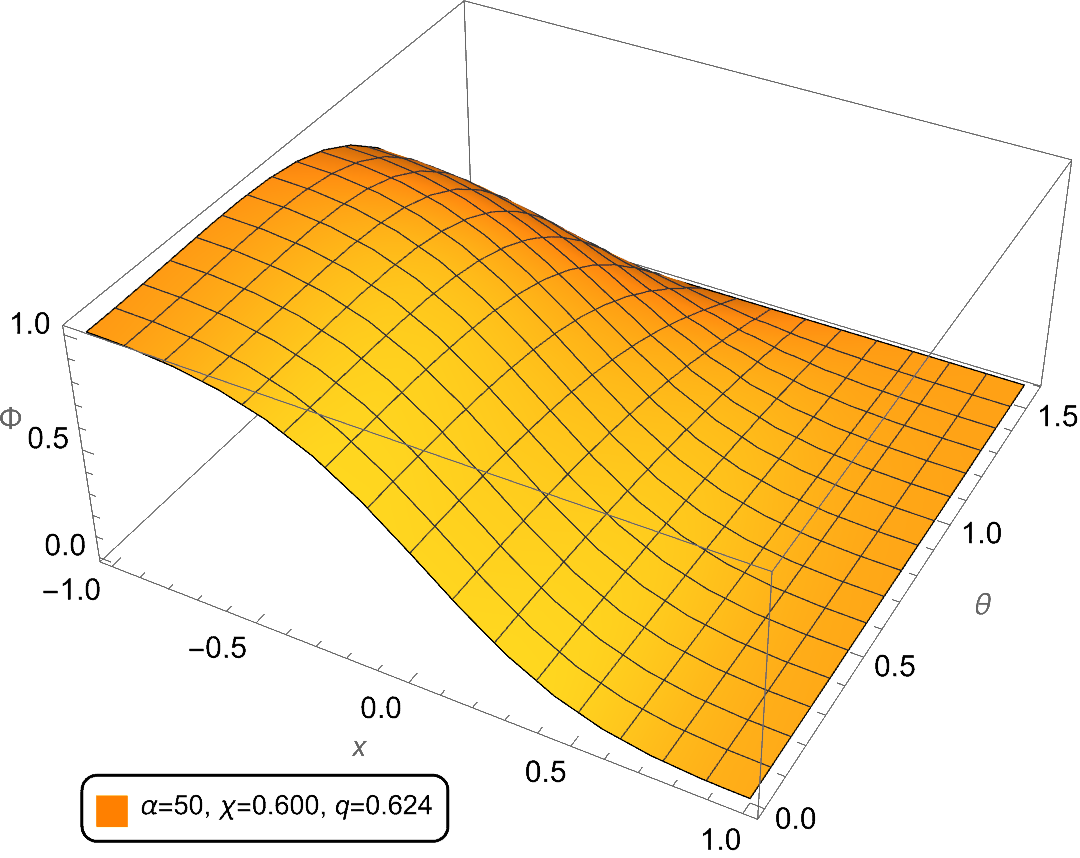} \includegraphics[scale=0.42]{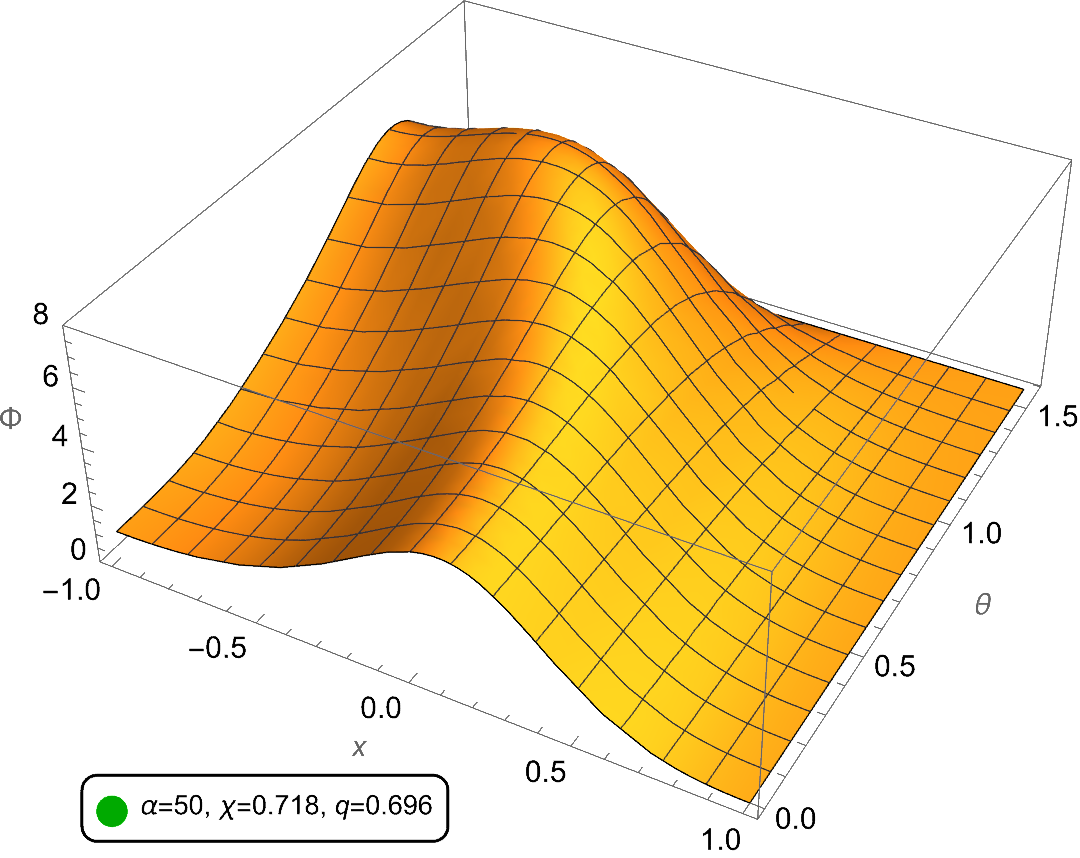}
\par\end{centering}
\caption{Existence domain and wave function of excited scalar clouds with
$\left(  n,l\right)  =\left(  0,1\right)  $. \textbf{Upper-Left Panel:}
Existence domain of excited clouds in the $\left(  \alpha,\chi,q\right)  $
space. \textbf{Upper-Right Panel:} Existence line for $\alpha=5$, $10$, $25$,
$50$ and $100$ in the $\left(  \chi,q\right)  $ space. While the lines for
$\alpha=5$ and $10$ resemble the fundamental case, those for $\alpha=25$, $50$
and $100$ increase as $\chi$ grows. \textbf{Middle Row}: Wave function of
three representative excited clouds with $\alpha=5$, vanishing at $\theta
=\pi/2$ due to odd parity. \textbf{Lower Row}: Wave function of three
representative excited clouds with $\alpha=50$. Black hole rotation induces a
pronounced bulge in scalar clouds, intensifying near the extremal limit.}%
\label{fig:Exwaveform01}%
\end{figure}

Fig. \ref{fig:Exwaveform01} presents the existence domain and wave function of
scalar clouds with $\left(  n,l\right)  =\left(  0,1\right)  $. For small
$\alpha$ (e.g., $\alpha=5$, $10$), existence line behavior in the $\left(
\chi,q\right)  $ plane resembles that of fundamental and $\left(  n,l\right)
=\left(  1,0\right)  $ modes. However, for sufficiently large $\alpha$ (e.g.,
$\alpha=25$, $50$, $100$), while existence lines still approach and terminate
at the extremal line, they exhibit an upward trend with increasing $\chi$,
leading to $q_{\text{RN}}\left(  \alpha\right)  <q_{\text{ext}}\left(
\alpha\right)  $. The middle row displays wave functions for three
representative excited clouds on the $\alpha=5$ existence line, vanishing at
$\theta=\pi/2$ due to odd parity. As black hole spin increases, these clouds
concentrate towards the event horizon. The lower row presents wave functions
for three representative excited clouds with $\alpha=50$, also vanishing at
$\theta=\pi/2$. Interestingly, a bulge emerges in the scalar clouds with
increasing black hole rotation, becoming pronounced near the extremal limit.

\begin{figure}[ptb]
\begin{centering}
\includegraphics[scale=0.37]{axD} \includegraphics[scale=0.37]{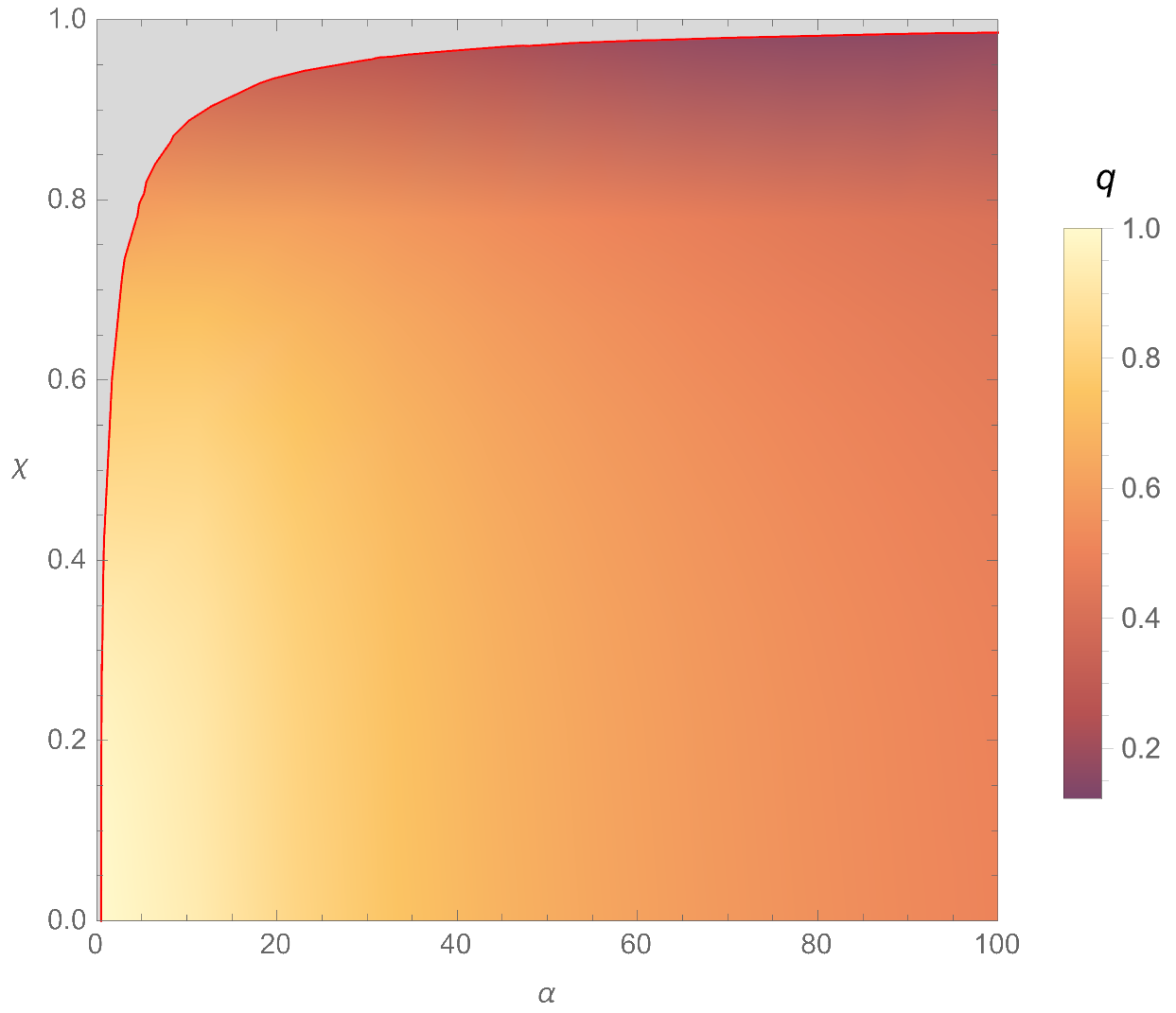}
\includegraphics[scale=0.37]{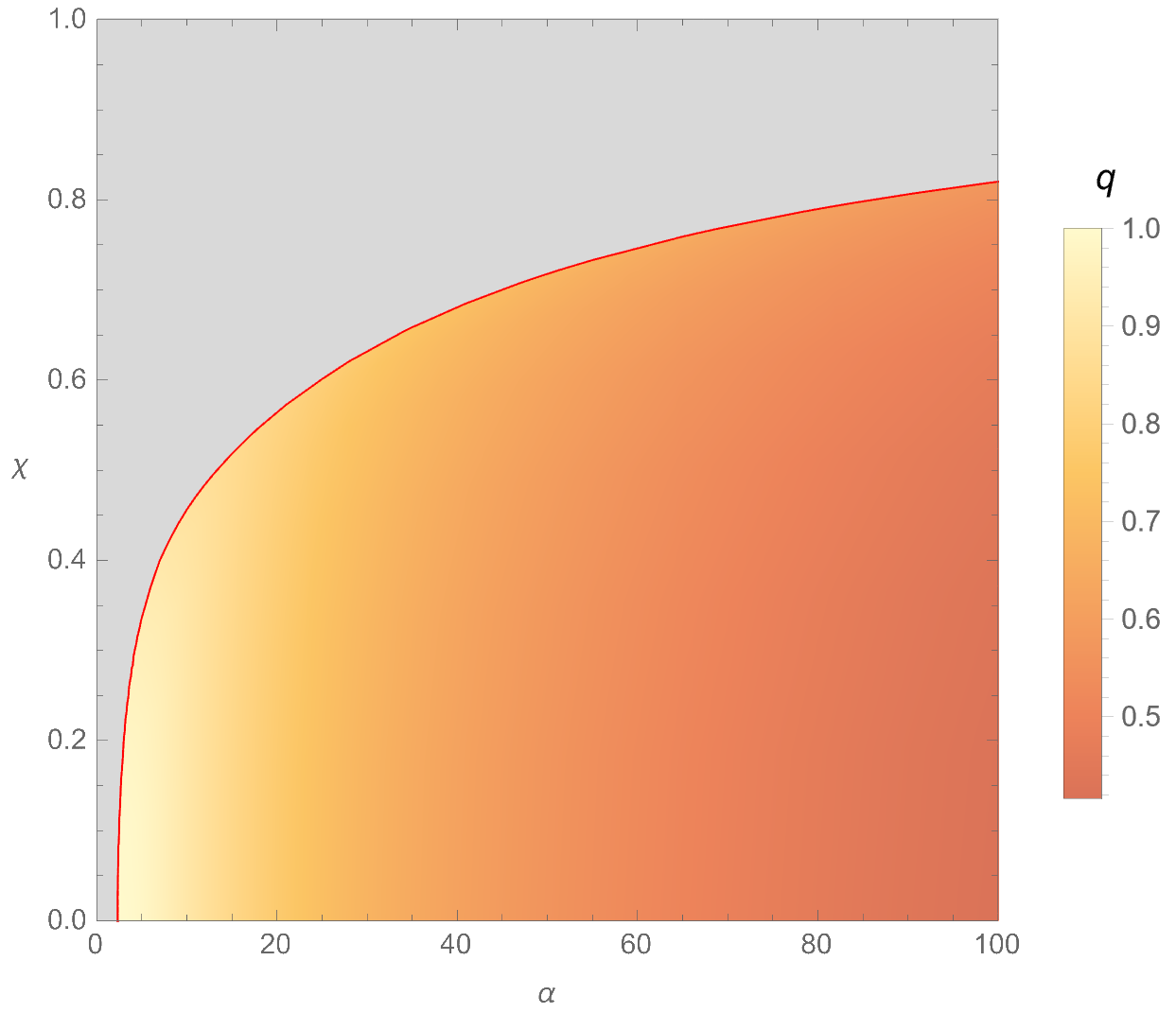} \includegraphics[scale=0.37]{aqD}
\includegraphics[scale=0.37]{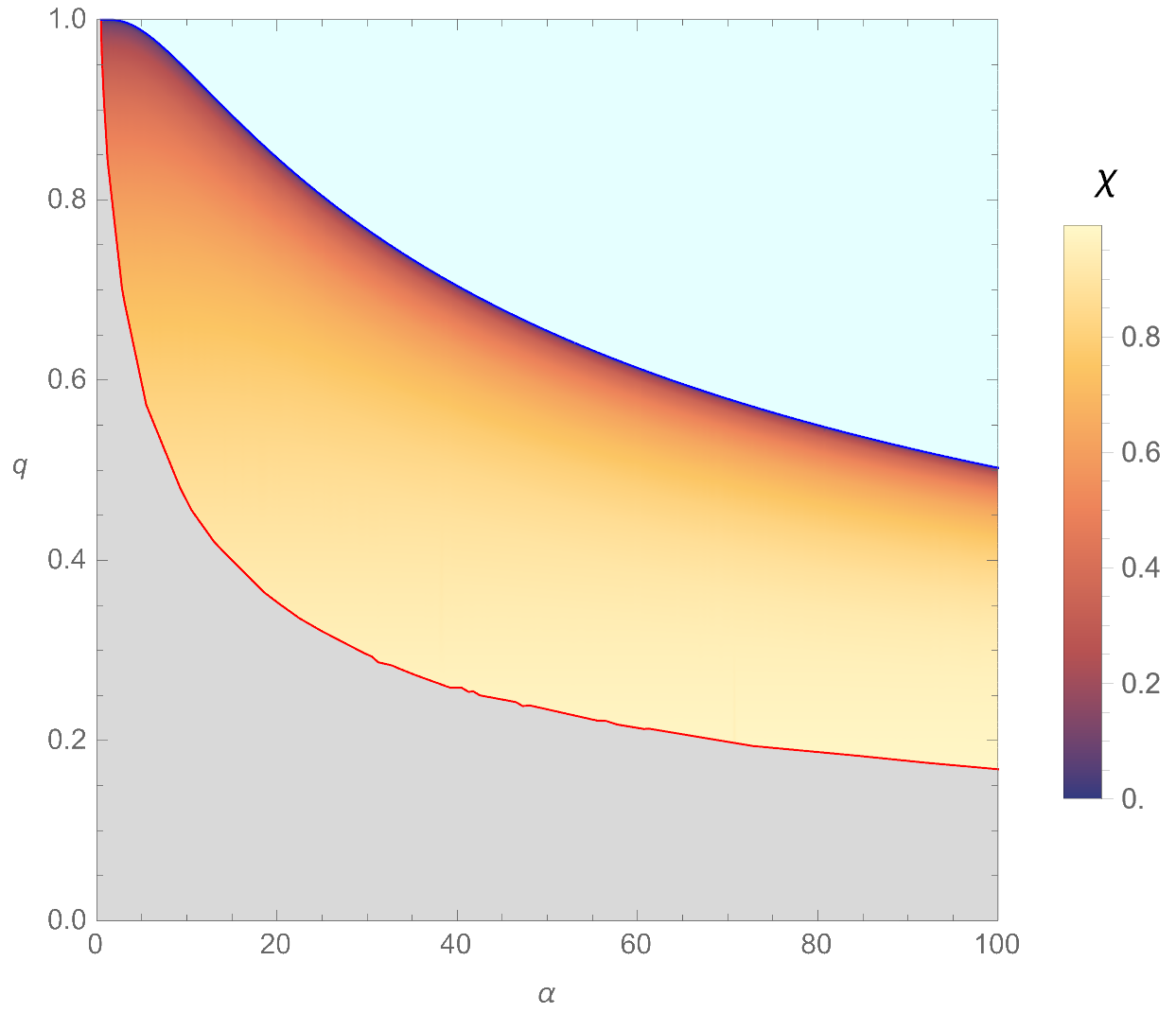} \includegraphics[scale=0.37]{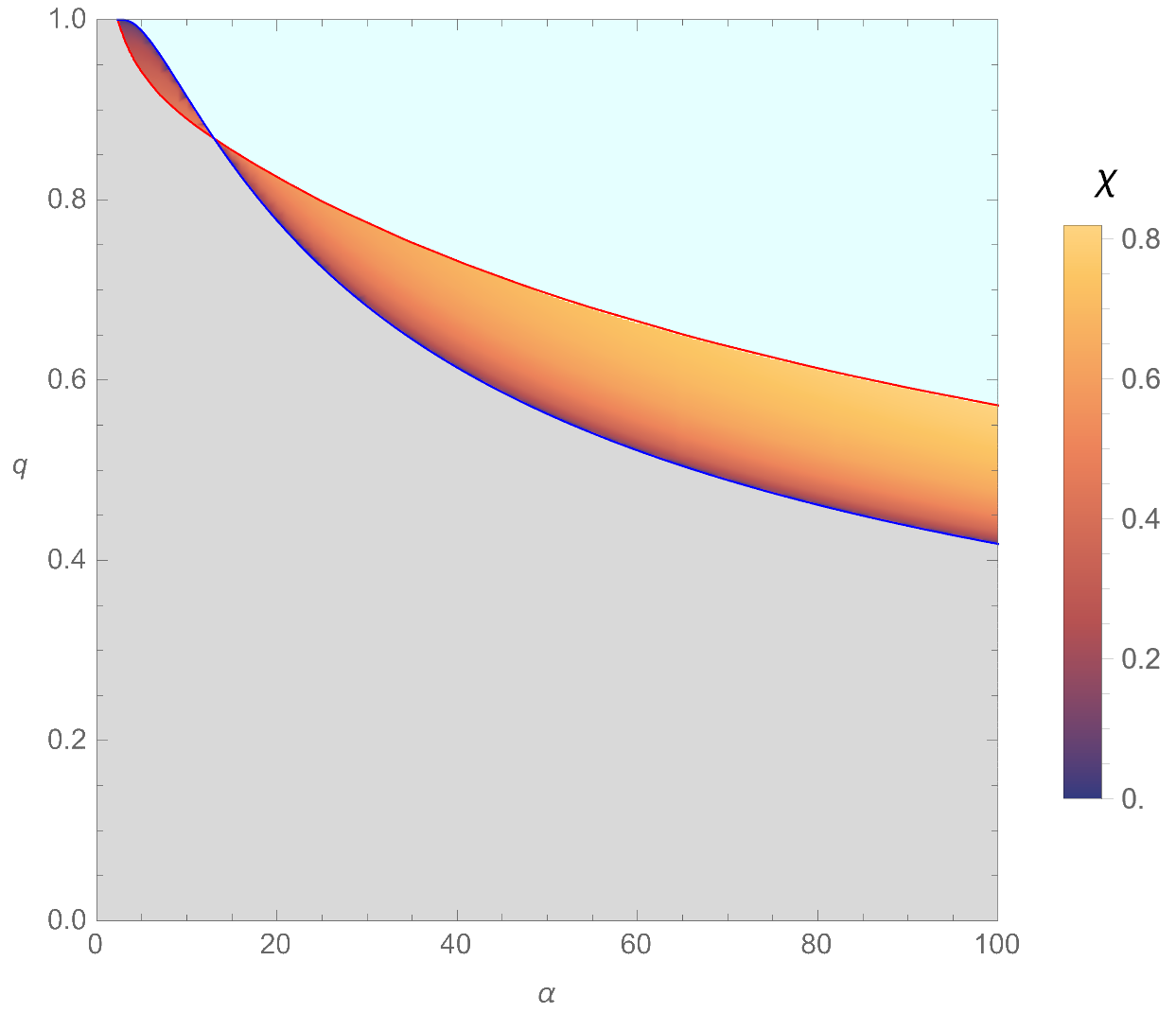}
\par\end{centering}
\caption{\textbf{Upper Row}: Existence domain of scalar clouds with $\left(
n,l\right)  =\left(  0,0\right)  $ (\textbf{Left Panel}), $\left(  n,l\right)
=\left(  1,0\right)  $ (\textbf{Middle Panel}) and $\left(  n,l\right)
=\left(  0,1\right)  $ (\textbf{Right Panel}) in the $\left(  \alpha
,\chi\right)  $ plane. Domains are bounded above by the $\chi_{\text{ext}%
}\left(  \alpha\right)  $ line (red lines). KN black holes supporting $l=1$
scalar clouds exhibit a lower maximum spin compared to the $l=0$ cases.
\textbf{Lower Row}: Existence domain of scalar clouds with $\left(
n,l\right)  =\left(  0,0\right)  $ (\textbf{Left Panel}), $\left(  n,l\right)
=\left(  1,0\right)  $ (\textbf{Middle Panel}) and $\left(  n,l\right)
=\left(  0,1\right)  $ (\textbf{Right Panel}) in the the $\left(
\alpha,q\right)  $ plane. Domains are bounded by the $\chi_{\text{ext}}\left(
\alpha\right)  $ line (red lines) and the $\chi_{\text{RN}}\left(
\alpha\right)  $ line (blue lines). Unlike $l=0$ cases, for large $\alpha$,
the upper and lower boundaries of $l=1$ cloud domains are determined by the
$\chi_{\text{ext}}\left(  \alpha\right)  $ and $\chi_{\text{RN}}\left(
\alpha\right)  $ lines, respectively. Cyan regions indicate parameter spaces
with excessively strong tachyonic instabilities for stationary scalar clouds.
Smaller cyan regions for excited clouds imply that the formation of excited
scalarized KN black holes requires stronger tachyonic instabilities.}%
\label{fig:Expara}%
\end{figure}

The upper and lower rows of Fig. \ref{fig:Expara} depict the existence domains
of two excited cloud modes in the $\left(  \alpha,\chi\right)  $ and $\left(
\alpha,q\right)  $ spaces, respectively. For comparison, the fundamental
cloud's existence domain is included in the left column. Gray regions indicate
insufficient tachyonic instabilities for scalar cloud formation, while cyan
regions exhibit excessively strong instabilities preventing stationary scalar
clouds. The upper row reveals an upper bound, $\chi_{\text{ext}}\left(
\alpha\right)  $, on KN black holes supporting scalar clouds. This bound is
lower for the $l=1$ mode compared to the $l=0$ modes, implying a requirement
for slower black hole spin to accommodate $l>0$ scalar clouds. In the $\left(
\alpha,q\right)  $ plane, $q_{\text{RN}}\left(  \alpha\right)  $ and
$q_{\text{ext}}\left(  \alpha\right)  $ define the upper and lower limits for
$l=0$ clouds. While this holds for small $\alpha$ values in the $l=1$ mode,
$q_{\text{RN}}\left(  \alpha\right)  $ and $q_{\text{ext}}\left(
\alpha\right)  $ serve as lower and upper boundaries for large $\alpha$.
Notably, gray regions for the excited modes are smaller than those of the
fundamental mode, indicating a stronger tachyonic instability threshold for
excited scalarized KN black holes.

\begin{figure}[ptb]
\includegraphics[scale=0.42]{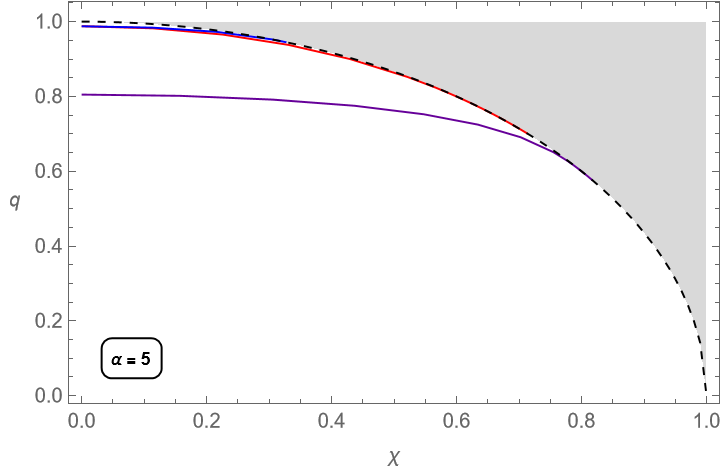} \includegraphics[scale=0.42]{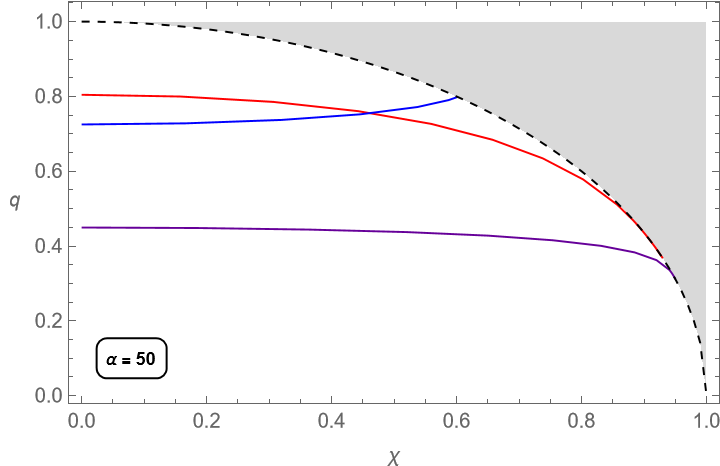}
\includegraphics[scale=0.42]{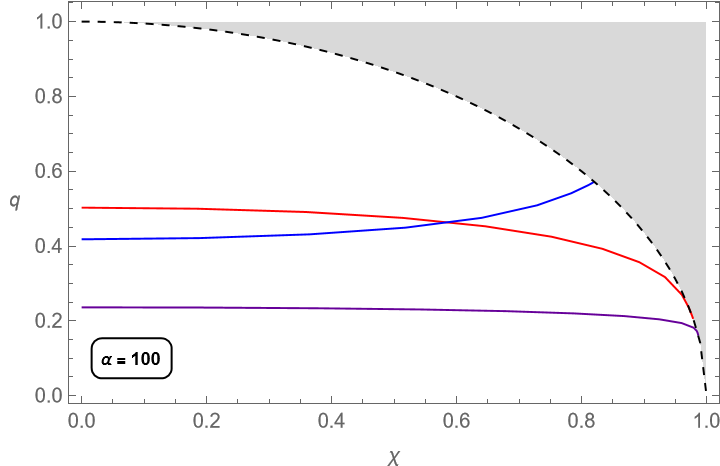}\caption{Existence line of fundamental and
excited clouds in the $\left(  \chi,q\right)  $ space for $\alpha=5$
\textbf{(Left Panel)}, $50$ \textbf{(Middle Panel) }and $100$ \textbf{(Right
Panel)}. In each panel, the clouds with $\left(  n,l\right)  =\left(
0,0\right)  $, $\left(  1,0\right)  $ and $\left(  0,1\right)  $ are shown by
purple, red and blue lines, respectively. The lowermost existence lines
correspond to fundamental clouds, indicating their greater susceptibility to
tachyonic instabilities.}%
\label{fig:Differ mode}%
\end{figure}

\begin{table}[ptb]%
\begin{tabular}
[c]%
{p{0.7cm}|p{1.4cm}p{1.4cm}p{1.4cm}|p{1.4cm}p{1.4cm}p{1.4cm}|p{1.4cm}p{1.4cm}p{1.4cm}}%
\hline
\multicolumn{1}{c|}{} & \multicolumn{9}{|c}{$q$}\\\cline{2-10}\cline{5-7}%
\multicolumn{1}{c|}{$\chi$} & \multicolumn{3}{|c|}{$\alpha=5$} &
\multicolumn{3}{|c|}{$\alpha=50$} & \multicolumn{3}{|c}{$\alpha=100$%
}\\\cline{2-10}\cline{5-7}%
\multicolumn{1}{c|}{} & \multicolumn{1}{|c}{$(0,0)$} &
\multicolumn{1}{c}{$(1,0)$} & \multicolumn{1}{c|}{$(0,1)$} &
\multicolumn{1}{|c}{$(0,0)$} & \multicolumn{1}{c}{$(1,0)$} &
\multicolumn{1}{c|}{$(0,1)$} & \multicolumn{1}{|c}{$(0,0)$} &
\multicolumn{1}{c}{$(1,0)$} & \multicolumn{1}{c}{$(0,1)$}\\\hline\hline
$0$ & $0.80502$ & $0.98787$ & $0.98762$ & $0.32858$ & $0.65457$ & $0.56234$ &
$0.23633$ & $0.50271$ & $0.41816$\\\hline
$0.1$ & $0.80367$ & $0.98357$ & $0.98481$ & $0.32833$ & $0.65338$ & $0.56342$
& $0.23617$ & $0.50194$ & $0.41905$\\\hline
$0.2$ & $0.79949$ & $0.97034$ & $0.97516$ & $0.32759$ & $0.64964$ & $0.56680$
& $0.23567$ & $0.49948$ & $0.42182$\\\hline
$0.3$ & $0.79209$ & $0.94725$ & $0.95382$ & $0.32628$ & $0.64283$ & $0.57291$
& $0.23480$ & $0.49498$ & $0.42678$\\\hline
$0.4$ & $0.78061$ & $0.91271$ & N/A & $0.32431$ & $0.63208$ & $0.58267$ &
$0.23348$ & $0.48783$ & $0.43460$\\\hline
$0.5$ & $0.76343$ & $0.86451$ & N/A & $0.32149$ & $0.61621$ & $0.59800$ &
$0.23160$ & $0.47724$ & $0.44652$\\\hline
$0.6$ & $0.73709$ & $0.79971$ & N/A & $0.31749$ & $0.59337$ & $0.62352$ &
$0.22895$ & $0.46201$ & $0.46511$\\\hline
$0.7$ & $0.69258$ & $0.71413$ & N/A & $0.31168$ & $0.56033$ & $0.67585$ &
$0.22512$ & $0.44021$ & $0.49612$\\\hline
$0.8$ & $0.59996$ & N/A & N/A & $0.30263$ & $0.50987$ & N/A & $0.21924$ &
$0.40761$ & $0.55372$\\\hline
$0.9$ & N/A & N/A & N/A & $0.28557$ & $0.41632$ & N/A & $0.20865$ & $0.35105$
& N/A\\\hline
\end{tabular}
\caption{Representative $q$ and $\chi$ values for the existence lines depicted
in FIG. \ref{fig:Differ mode}.}%
\label{tab:valueList}%
\end{table}

Fig. \ref{fig:Differ mode} illustrates the existence lines of fundamental and
excited clouds in the $\left(  \chi,q\right)  $ plane for $\alpha=5$, $50$ and
$100$. Representative $q$ and $\chi$ values for each existence line are
tabulated in Tab. \ref{tab:valueList}. For all $\alpha$, fundamental cloud
existence lines lie below those of excited clouds, indicating a requirement
for stronger tachyonic instabilities to form excited clouds. When the coupling
constant $\alpha$ is weak, $\left(  n,l\right)  =\left(  1,0\right)  $ excited
cloud existence lines primarily reside below those of $\left(  n,l\right)
=\left(  0,1\right)  $ clouds, suggesting a larger $q$ (and stronger tachyonic
instabilities) for sustaining $l=1$ clouds. Conversely, for strong $\alpha$,
slowly rotating black holes require less charge to support $\left(
n,l\right)  =\left(  0,1\right)  $ clouds, whereas higher charges are
necessary for these clouds when black hole rotation accelerates.

\section{Conclusions}

\label{sec:Conclusion}

In this work, we have explored the existence and characteristics of both
fundamental and excited scalar clouds in KN black holes. Through a detailed
analysis of the parameter space, we identified the regions where these clouds
can form and examined how varying the coupling constant $\alpha$, black hole
spin $\chi$ and charge $q$ influences the formation of these clouds.

For fundamental scalar clouds, our results underscore the influence of the
coupling constant $\alpha$ and black hole parameters on the cloud's existence.
As expected, larger coupling constants $\alpha$ facilitate the formation of
these clouds by reducing the required charge. Furthermore, the black hole spin
imposes an upper limit on the existence of these clouds, with faster spins
leading to a higher charge threshold. Rapidly rotating black holes exhibit a
concentration of scalar clouds near the event horizon. In contrast, excited
scalar clouds require stronger tachyonic instabilities, as indicated by their
existence lines, which consistently lie above those of fundamental clouds.
These excited clouds also display unique wave function characteristics, such
as valleys that shift toward the event horizon and the emergence of pronounced
bulges as the spin increases.

These findings deepen our understanding of scalar cloud dynamics around KN
black holes and suggest promising avenues for future research. Excited scalar
clouds offer a potential starting point for investigating scalarized KN black
holes in excited states. Given the broader applicability of the spectral
method, extending our analysis to other rotating black hole solutions is a
natural next step. Finally, exploring the astrophysical implications of scalar
clouds, particularly within the context of gravitational wave astronomy,
presents a compelling research direction.

\begin{acknowledgments}
We are grateful to Yiqian Chen for useful discussions and valuable comments.
This work is supported in part by NSFC (Grant No. 12105191, 11947225 and 11875196).
\end{acknowledgments}

\bibliographystyle{unsrturl}
\bibliography{ref}

\end{document}